\documentclass[fleqn,usenatbib]{mnras}

\usepackage{newtxtext,newtxmath}
\usepackage[T1]{fontenc}
\usepackage{graphicx}
\usepackage{amsmath}
\usepackage{xcolor}
\usepackage{lastpage}
\usepackage{orcidlink}

\DeclareRobustCommand{\VAN}[3]{#2}
\let\VANthebibliography\thebibliography
\def\thebibliography{\DeclareRobustCommand{\VAN}[3]{##3}\VANthebibliography}

\title[Cores and Rings of Hoag-type Galaxies]{Hoag-type Galaxies: A Photometric Study of the Cores and Rings of Thirty Systems from the DESI Legacy Survey}

\author[J. Yang \& C. Li]{
Jingshuo Yang\orcidlink{0009-0008-2209-2803}$^{1}$\thanks{E-mail: yjs25@mails.tsinghua.edu.cn}
and Cheng Li\orcidlink{0000-0002-8711-8970}$^{1}$\thanks{E-mail: cli2015@tsinghua.edu.cn}
\\
$^{1}$Department of Astronomy, Tsinghua University, Beijing 100084, China
}

\date{}
\pubyear{\the\year{}}

\begin{document}
\label{firstpage}
\pagerange{\pageref{firstpage}--\pageref{LastPage}}
\maketitle

\begin{abstract}
Hoag-type galaxies contain a smooth central core surrounded by a detached outer ring, but their rarity has limited previous studies mainly to individual systems. We identify thirty galaxies with apparent Hoag-type morphology in DESI imaging by visually inspecting two machine-learning ring-galaxy catalogues and analyse their core and ring properties using DESI Legacy Imaging Surveys $g$, $r$, and $z$ images. We measure the rings after subtracting two-dimensional core models and compare the cores with matched SDSS galaxies. The cores lie predominantly on the red sequence, whereas the rings span a broader colour range extending toward the blue cloud. The cores broadly follow early-type mass scaling relations, but their mass-matched distributions tend towards slightly bluer colours, smaller sizes, higher stellar-mass surface densities, and lower concentrations and S\'ersic indices. Of the 16 systems with SDSS group information, 14 (87.5 per cent) are centrals, compared with $\sim64\%$ in the mass-matched early-type control sample. The halo-mass and large-scale overdensity distributions become much more similar after central--satellite status is also matched. Ring stellar-mass surface density and normalized width decrease with increasing core-plus-inner-halo to ring mass ratio. Most core $R_{90}$ values lie inside the outermost stable circular orbit of a simplified core-plus-ring model, while the rings frequently extend across its unstable interval. The slightly bluer core colours, lower concentrations, and lower S\'ersic indices suggest a possible secular contribution, but the compact, dense cores imply more complex structural assembly than a simple diffuse-pseudobulge origin. The high central fraction may favour gas retention and ring survival by reducing satellite-specific environmental processing. The diversity of ring properties nevertheless leaves room for multiple formation pathways.
\end{abstract}


\section{Introduction}

Hoag's Object, discovered by \citet{Hoag1950}, is the prototype of Hoag-type galaxies. It is a peculiar ring galaxy consisting of a smooth, nearly round core surrounded by a detached blue ring. Although rings occur in diverse galactic structures, from resonance and collisional rings to polar systems, Hoag’s Object stands out for the spatial separation of its two luminous components: a pronounced low-surface-brightness gap isolates the ring from the core. Spectroscopy established that the core and ring are physically associated, excluding early planetary-nebula and gravitational-lens interpretations \citep{OConnell1974,Schweizer1987}. The core follows an $r^{1/4}$ profile and has the size and velocity dispersion of a normal spheroid; its old stellar population, red colour gradient, twisted isophotes, and high angular momentum indicate a mildly triaxial elliptical galaxy without a detectable bar or disc \citep{Schweizer1987,Finkelman2011Hoag}. The ring, by contrast, is comparably luminous but younger, knotty, and rich in H$\alpha$ emission. It forms a quasi-spiral pattern within a regularly rotating H\,{\sc i} structure that extends to about twice the optical-ring diameter \citep{Finkelman2011Hoag,Brosch2013}. The regular gas kinematics and lack of a nearby massive companion exclude a recent strong interaction, but do not rule out an older event \citep{Brosch2013}.

Only a few proposed analogues have been studied in detail. NGC 6028, UGC 4599, and PGC 1000714 have all been identified from their apparently smooth central components and surrounding ring-like structures, and none has an obvious intruder for a classical collisional ring. Their detailed morphologies nevertheless differ. NGC 6028 has an oval, weak-bar-like core, detectable light in the gap, and a diffuse outer ring \citep{Wakamatsu1990}. UGC 4599 was initially described as having an $r^{1/4}$ central body and an extremely extended H\,{\sc i} disc \citep{Finkelman2011UGC}; deeper imaging subsequently revealed spiral structure connecting the centre to the apparent ring and a giant low-surface-brightness disc, challenging its Hoag-type classification \citep{Mosenkov2023}. PGC 1000714 more closely resembles Hoag's Object in having an elliptical core without a bar or stellar disc and a complete outer ring, but it also contains a redder inner ring; the three components have distinct stellar populations \citep{MutluPakdil2017}. These systems therefore differ in both their underlying structures and stellar populations, illustrating the diversity of apparent Hoag-type morphologies and the need to establish whether their outer structures are genuinely detached.

Three main formation pathways have been proposed. In bar-driven secular evolution, a bar moves disc material toward an outer resonance and later dissolves \citep{Brosch1985}. Simulations can thereby produce a long-lived circular ring and depleted inner region after the bar disappears \citep{Freeman2010}; however, this mechanism requires a pre-existing disc and does not readily explain the elliptical-like core of Hoag's Object. Alternatively, a merger or external-accretion event can deposit high-angular-momentum gas around a spheroid, where it settles and forms stars. Events several Gyr ago have been proposed for Hoag's Object and UGC 4599, while more recent accretion may explain PGC 1000714 \citep{Schweizer1987,Finkelman2011UGC,MutluPakdil2017}. Such old events could have lost their tidal signatures, but are consequently difficult to verify, and the regular H\,{\sc i} disc of Hoag's Object excludes a recent strong interaction \citep{Brosch2013}. A third possibility is prolonged cold accretion from the intergalactic medium, which builds a low-density H\,{\sc i} disc around an old elliptical; star formation then occurs only in density enhancements induced by its triaxial potential \citep{Finkelman2011Hoag}. This explains the old core, extended gas, and isolation without a disruptive merger, but the accreted gas and its expected chemical signature remain unconfirmed, and the mechanism does not itself explain the gap.

A complementary dynamical model addresses the gap rather than the origin of the ring. Approximating a homogeneous torus by a thin ring of the same mass \citep{Bannikova2011}, \citet{Bannikova2018} represented a Hoag-like system as a central point mass plus a massive circular ring. Their competing gravitational forces define a Lagrangian circle (LC), while stable circular orbits around the core end at an outermost stable circular orbit (OSCO). Circular motion is unstable between the OSCO and LC, so depletion from this interval could create or maintain the gap; both radii depend on the core-to-ring mass ratio. The model is, however, highly idealized and omits the extended core, dark-matter halo, gas dynamics, non-circular motions, and ring formation history. It therefore supplies a useful stability scale rather than a complete formation scenario.

The rarity of Hoag-type galaxies and the uncertainty among these formation pathways have restricted previous work largely to individual systems. Wide-area surveys and machine-learning catalogues of ring galaxies now make a statistical study possible. We visually select thirty Hoag-type galaxies from SDSS- and DESI-based catalogues and perform homogeneous core--ring photometry using DESI Legacy Imaging Surveys $g$, $r$, and $z$ images. We compare the component colours, luminosities, structures, surface densities, and stellar masses, examine the cores relative to matched SDSS galaxies and their environments, and test the observed radial structure against the model of \citet{Bannikova2018}.

Throughout this paper we adopt a flat $\Lambda$CDM cosmology with $H_0=70$ km s$^{-1}$ Mpc$^{-1}$ and $\Omega_{\rm m}=0.3$. This paper is organized as follows. Section~\ref{sec:sample_data} describes the identification, imaging data, and photometric measurements of the Hoag-type galaxies. Sections~\ref{sec:results} and \ref{sec:ring_structure} present the main results, Section~\ref{sec:discussion} discusses their implications, and Section~\ref{sec:summary} summarizes our conclusions.

\section{Identification and photometry}
\label{sec:sample_data}

\subsection{Identification}
\label{subsec:sample}

Our parent sample is drawn from two recently constructed catalogues of ring galaxies selected with machine-learning methods. \citet{Abraham2025} applied a modified AlexNet classifier to SDSS DR18 colour-composite images. Their prediction sample contained 29,420 relatively nearby and bright SDSS galaxies selected with criteria similar to their training sample, and 4,855 of these systems had ring-like structures with prediction confidence above 90 per cent.

\citet{Zhang2025} searched a much larger DESI Legacy Imaging Surveys DR9 sample using a two-stage binary classifier based on the Swin Transformer architecture. Their first classifier selected possible ring-galaxy candidates from 573,668 DR9 galaxy images with spectroscopic redshifts $z_{\rm spec}=0.01$--0.20 and $m_r<17.5$, and their second classifier was designed to remove common contaminants, especially spiral and barred spiral galaxies. After combining the candidate lists, visual inspection confirmed 12,196 true ring galaxies from 18,802 unique candidates. The authors then removed objects already present in their training sample and in previous catalogues, leaving 8,052 newly discovered ring galaxies.

\begin{figure*}
    \includegraphics[width=\textwidth]{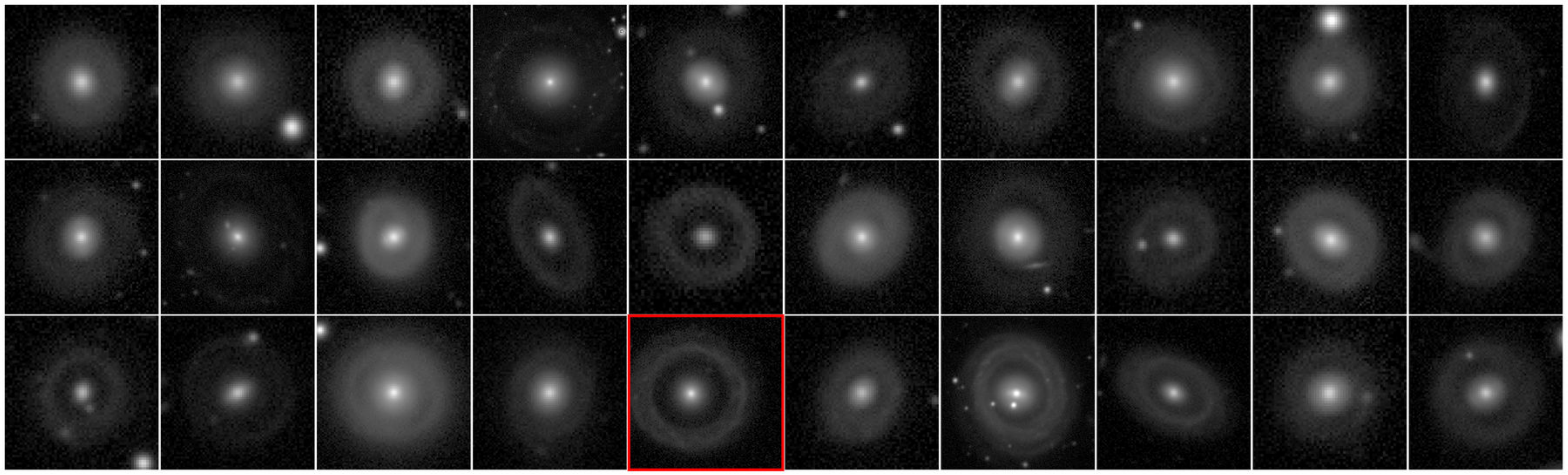}
    \caption{Images of the thirty Hoag-type galaxies selected for this work. Hoag's Object is marked by the red frame.}
    \label{fig:sample_examples}
\end{figure*}

We combined the SDSS- and DESI-based ring-galaxy catalogues and visually inspected the resulting candidates to search for Hoag-type systems. The visual selection focused on the morphology that distinguishes Hoag-type galaxies from ordinary resonance rings and collisional ring galaxies: a smooth central core, no obvious strong bar, a detached outer ring, and a visible gap between the core and ring. Systems with clear spiral arms connecting the core and ring, strongly distorted collisional morphology, or ordinary barred-ring structure were excluded. This procedure yielded thirty Hoag-type galaxies. The previously studied systems Hoag's Object, NGC 6028, UGC 4599, and PGC 1000714 are all recovered by this selection. Their inclusion reflects their appearance at the depth of the DESI images; the classification of UGC 4599 in deeper observations and its implications for sample purity are discussed in Section~\ref{subsec:sample_purity}. Their images are shown in Figure~\ref{fig:sample_examples}, and their coordinates and redshifts are listed together with the core properties in Table~\ref{tab:core_properties}.

\subsection{Photometric data}
\label{subsec:data}

All imaging data used for the photometric measurements in this work are taken from Data Release 9 of the DESI Legacy Imaging Surveys\footnote{\url{https://www.legacysurvey.org/dr9/description/}} \citep{Dey2019}. The Legacy Surveys were designed to provide deep, wide-field imaging for DESI target selection and for broader studies of galaxies over a large extragalactic footprint. They combine three public imaging programmes: the DECam Legacy Survey (DECaLS), the Beijing-Arizona Sky Survey (BASS), and the Mayall $z$-band Legacy Survey (MzLS). Together these surveys image approximately 14,000 deg$^2$ of sky in the optical $g$, $r$, and $z$ bands, with additional mid-infrared photometry from WISE/NEOWISE. In the northern part of the footprint, the optical imaging is mainly provided by BASS in $g$ and $r$ and by MzLS in $z$; in the southern part, it is mainly provided by DECaLS in all three optical bands \citep{Dey2019}.

DR9 is the ninth public release of these images and catalogues. It builds on the earlier full-footprint releases by improving the reduction and processing of the Legacy Surveys imaging, and provides coadded images and Tractor catalogues in sky ``bricks'' of approximately $0.25^\circ \times 0.25^\circ$. The native optical image stacks are tangent-plane coadds with a pixel scale of 0.262 arcsec pixel$^{-1}$, and the Legacy Survey viewer can return FITS cutouts from the combined DR9 layer. In this work, the science cutouts for all thirty galaxies were downloaded from the Legacy Survey viewer using \texttt{layer=ls-dr9}. The cutouts have a pixel scale of 0.5 arcsec pixel$^{-1}$, except for the larger, nearby galaxy UGC 4599, for which we use a 1.0 arcsec pixel$^{-1}$ cutout. The Legacy Surveys photometry is reported on the AB natural system, with fluxes stored in nanomaggies and magnitudes given by $m=22.5-2.5\log_{10}({\rm flux})$. When integrating a surface-brightness profile, we therefore multiply the annular areas measured in cutout pixels by $(p/0.262)^2$, where $p$ is the cutout pixel scale in arcsec pixel$^{-1}$.

\subsection{Photometry}
\label{sec:decomposition}

The main challenge in Hoag-type photometry is to separate the light of the central core from that of the detached ring. Most previous studies of individual Hoag-type systems measured the ring through aperture photometry with the central region masked or excluded \citep[e.g.][]{Schweizer1987,Finkelman2011Hoag,Finkelman2011UGC}. In such measurements, diffuse core light underlying the ring may contribute to the measured ring flux. However, core-model subtraction has also been used: \citet{MutluPakdil2017} measured the rings of PGC 1000714 in residual images after subtracting a GALFIT core model. Here we apply a consistent core-model subtraction procedure across the full sample. We first mask contaminating sources and determine the local sky background (\S\ref{subsec:mask_background}). We then mask the ring, extract the one-dimensional surface-brightness profile of the core, and fit it with an appropriate radial model. Because the cores have relatively simple and approximately centre-symmetric structures, the fitted profile can be combined with the measured isophotal geometry to construct a two-dimensional core model (\S\ref{subsec:core_profiles}). Finally, we subtract the core model from each image and measure the ring profile from the resulting residual image (\S\ref{subsec:ring_profiles}). This procedure removes the diffuse core light underlying the ring while preventing ring light from affecting the core fit, providing consistently separated photometry and radial profiles for the two components.

\subsubsection{Masking and background subtraction}
\label{subsec:mask_background}

\begin{figure*}
    \includegraphics[width=\textwidth]{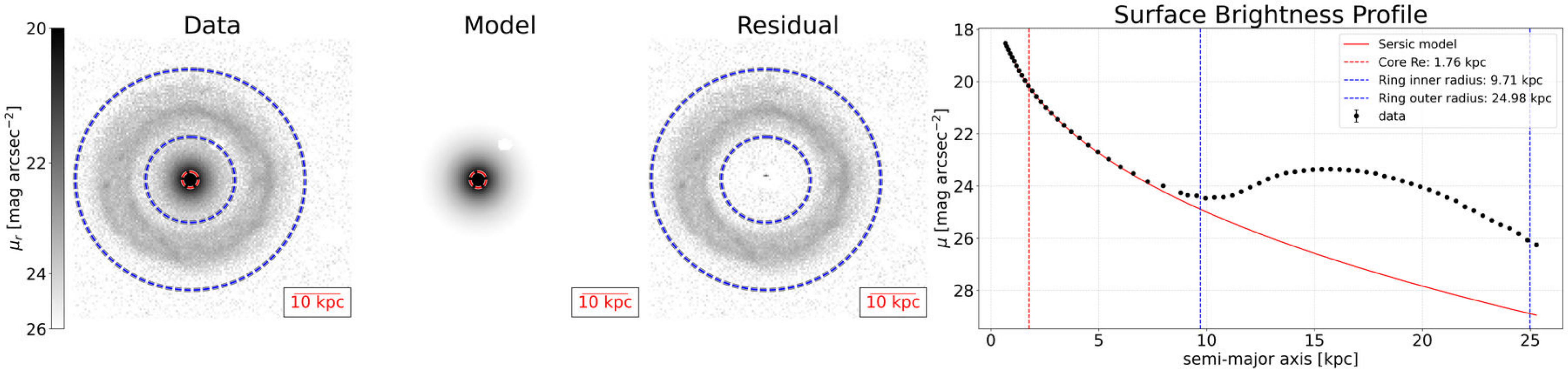}
    \caption{Example of the photometric decomposition procedure for Hoag's Object. From left to right: masked science image, two-dimensional core model, residual image after subtracting the core model, and the measured surface-brightness profile with a S\'ersic fit to the core. The red dashed line marks the core effective radius $R_{\rm e}$, and the blue dashed lines mark the inner and outer ring radii.}
    \label{fig:decomposition_example}
\end{figure*}

The first step is to remove light that does not belong to the target galaxy. The initial source mask was generated with SExtractor \citep{BertinArnouts1996}. We run SExtractor on the stacked $g+r+z$ image with the parameter \texttt{DETECT\_THRESH} set to 0.5, 1, 2, and 3 and combine the resulting segmentation maps. The segmentation label containing the target centre is excluded from the contaminant mask, and additional object-specific labels identified by visual inspection as parts of the target are also retained. For contaminants detected at more than one threshold, the lowest-threshold segmentation that can be separated reliably from the target is used to retain their faint outskirts. Motivated by studies showing that insufficient masks around extended sources can lead to sky oversubtraction \citep{Ji2018,Kelvin2023}, the combined mask is expanded with a fixed $5\times5$ cross-shaped convolution kernel. Finally, each image is inspected visually and additional DS9 regions are used to remove residual contaminants or artefacts that are not captured by the automatic segmentation. The left panel of Figure~\ref{fig:decomposition_example} shows Hoag's Object after this masking step.

After masking, we estimate and subtract the local sky background independently in each band. The background region outside the outer ring is divided into four concentric annuli and four quadrants, giving $K=16$ subregions. Within each subregion $j$, the unmasked pixels are iteratively clipped about their median using the median absolute deviation. We denote the clipped median, standard deviation, and number of retained pixels by $m_j$, $s_j$, and $N_j$, respectively. The adopted local background is the unweighted mean of the $K$ usable regional medians,
\begin{equation}
    I_{\rm sky}=\frac{1}{K}\sum_{j=1}^{K}m_j .
\end{equation}

The resampled cutouts contain correlated pixel noise. We estimate an empirical correlation factor $C_{\rm corr}$ by comparing the observed dispersion of blank-sky block means, using block sizes of 4, 8, and 16 pixels, with the dispersion expected for independent pixels. Each regional median is removed before this calculation so that large-scale background structure is not interpreted as pixel correlation. For a Gaussian distribution, the estimated standard error of regional median $m_j$ is
\begin{equation}
    e_j=1.2533\,C_{\rm corr}\frac{s_j}{\sqrt{N_j}}.
\end{equation}
Motivated by the separation of high- and low-frequency background variations in \citet{GilDePazMadore2005}, we define
\begin{align}
    \sigma_{\rm sky,HF}^{2} &= \frac{1}{K^2}\sum_{j=1}^{K}e_j^2,\\
    \sigma_{\rm sky,LF}^{2} &= \max\left[s_m^2-\frac{1}{K}\sum_{j=1}^{K}e_j^2,0\right],\\
    \Delta I_{\rm sky}^{2} &= \sigma_{\rm sky,HF}^{2}+\sigma_{\rm sky,LF}^{2},
    \label{eq:sky_uncertainty}
\end{align}
where $s_m$ is the sample standard deviation of the $K$ regional medians. The first term is the finite-sampling variance of their mean, while the second retains residual region-to-region variation as a coherent uncertainty in the sky level beneath the galaxy. The same $C_{\rm corr}$ is also applied to the random isophotal uncertainties described below; no additional pixel-noise term is added to them.

\subsubsection{Core profiles and two-dimensional models}
\label{subsec:core_profiles}
\label{subsec:core_fitting}
\label{subsec:core_model}

The core profile is measured after the ring and contaminating sources have been masked and the local sky background has been subtracted. We use the core measurement to build an explicit model that is later subtracted before the ring is measured.

Following the standard technique of fitting the intensity distribution along nested ellipses \citep{Jedrzejewski1987}, we extract the core light profile with elliptical isophotes using the \texttt{photutils.isophote.Ellipse}\footnote{\url{https://photutils.readthedocs.io/en/stable/reference/isophote_api.html}} implementation \citep{Bradley2024Photutils}. The isophotal geometry is determined from the sky-subtracted, masked $g+r+z$ image. The core centre and position angle are held fixed and the ellipticity is allowed to vary with semi-major axis. The ring mask is applied throughout so that ring light does not set the core geometry or intensity. The resulting isophote list provides the semi-major axis, centre, ellipticity, and position angle for each fitted ellipse. To obtain matched multi-band profiles, the same isophotal geometry is then applied to the individual $g$, $r$, and $z$ images. Intensities are sampled with the median integration mode.

We then fit three candidate models to each one-dimensional core profile: a single S\'ersic profile \citep{Sersic1968,GrahamDriver2005}, a continuous piecewise model with an inner S\'ersic and an outer exponential profile, and a continuous double-S\'ersic model. These piecewise models are intended to capture profiles with systematic slope changes, which may indicate weak bars, lenses, or other extended central structures rather than a purely smooth spheroid. For each adopted model we sample $I_{\rm model}(R;\boldsymbol\theta)+\delta_{\rm sky}$ with the affine-invariant ensemble sampler \texttt{emcee} \citep{ForemanMackey2013}. Here $\delta_{\rm sky}$ is one background offset shared by all radii in a given image, with Gaussian prior $\delta_{\rm sky}\sim\mathcal{N}(0,\Delta I_{\rm sky}^{2})$. The Gaussian likelihood uses $C_{\rm corr}\,\sigma_{\rm iso,core}(R)$ as the random uncertainty of each isophote. The model parameters have bounded uniform priors over the adopted fitting ranges. Parameter estimates are posterior medians, with half the 16th-to-84th percentile interval reported as the statistical uncertainty. The core-plus-sky uncertainty at radius $R$ is the standard deviation of $I_{\rm model}(R;\boldsymbol\theta^{(d)})+\delta_{\rm sky}^{(d)}$ over the joint posterior draws, denoted by $\sigma_{\rm core}(R)$. This retains covariance among the fitted parameters and the shared background offset while conditioning on the selected model family. The final model is chosen by combining the Bayesian Information Criterion (BIC) ranking with visual inspection of the residuals. With this procedure, 16 of the 30 galaxies are best described by a single S\'ersic profile, ten by the S\'ersic-plus-exponential piecewise model, and four by the double-S\'ersic piecewise model. The right panel of Figure~\ref{fig:decomposition_example} shows the measured profile and adopted core model for Hoag's Object.

The one-dimensional fit is converted into a two-dimensional model using the measured isophotal geometry. At each fitted semi-major axis, the measured intensity is replaced with that predicted by the selected radial model while the centre, ellipticity, and position angle are retained. The final isophotes are extended outward to estimate the diffuse core contribution at the ring radius. The central $4\times4$ pixel block of the two-dimensional subtraction model is replaced by the observed image values to avoid numerical artefacts at the innermost isophotes. The second panel of Figure~\ref{fig:decomposition_example} shows the resulting core model for Hoag's Object. 

\subsubsection{Ring profiles from core-subtracted residual images}
\label{subsec:ring_profiles}

For each band, we subtract the adopted sky level and the two-dimensional core model from the original science image. The result is a residual image in which the detached ring is the dominant large-scale component, as shown in the third panel of Figure~\ref{fig:decomposition_example}. We sum the $g$, $r$, and $z$ residual images and use the manually defined ring annulus to specify the region over which flux-weighted moments estimate the ring centre, ellipticity, and position angle. The resulting geometry is held fixed while median intensities are sampled from the image centre to the boundary in each of the three residual images.

This residual-based measurement is designed to avoid assigning the extrapolated core light to the ring. For posterior draw $d$, the coherent correction to the nominal residual profile is
\begin{equation}
    c^{(d)}(R)=\delta_{\rm sky}^{(d)}+
    I_{\rm core}(R;\boldsymbol\theta^{(d)})-
    I_{\rm core}(R;\widetilde{\boldsymbol\theta}),
\end{equation}
where $\widetilde{\boldsymbol\theta}$ is the vector of median fitted parameters used to construct the nominal core model. The marginal uncertainty used to display and detect the ring at a given radius is
\begin{equation}
    \sigma_{\rm ring}^{2}(R)=
    \left[C_{\rm corr}\,\sigma_{\rm iso,ring}(R)\right]^2+
    {\rm Var}_{d}\left[c^{(d)}(R)\right].
\end{equation}
The first term is random isophotal noise and the second is the coherent uncertainty from the joint core-plus-sky posterior. Together with the core profiles described above, these ring profiles form the basis for the subsequent measurements of the photometric and structural properties of the galaxies.

\subsection{Derived photometric quantities}
\label{subsec:derived_quantities}

For each galaxy we measure the integrated photometric quantities of the core and ring separately from the profiles described in Section~\ref{sec:decomposition}. To ensure a consistent comparison with the SDSS samples, we define the core integration aperture using the Petrosian ratio in the $r$ band. This ratio compares the mean surface brightness in the elliptical annulus from $0.8R$ to $1.25R$ with that within $R$. The Petrosian radius $R_{\rm P}$ is determined where the ratio falls to 0.2, and the core flux is integrated within $2R_{\rm P}$, following the SDSS Petrosian convention of \citet{Blanton2001}. This aperture is held fixed when integrating the posterior core profiles in all three bands. For every posterior draw, we determine $R_{\rm e}$ and $R_{90}$ from the radii enclosing 50 and 90 per cent of the integrated core light within this aperture.

For the ring, we identify the contiguous run of at least two $r$-band samples with $I_{\rm ring}/\sigma_{\rm ring}>2$ that contains the strongest peak inside the manually defined ring annulus. The threshold crossings are linearly interpolated to define the inner and outer radii. This annulus is also held fixed when integrating the posterior ring profiles in all three bands. Reported fluxes and their statistical uncertainties are the medians and half the 16th-to-84th percentile intervals within this annulus.

The annular integration weights follow the measured ellipticity and are multiplied by $(p/0.262)^2$ to account for the cutout pixel scale. Absolute magnitudes are calculated from the integrated fluxes and the luminosity distance in the adopted flat cosmology, and are corrected for Galactic extinction. No $K$-correction is applied. Colours are measured from the corresponding component magnitudes. The core concentration is defined as $C=R_{90}/R_{\rm e}$. The reported core S\'ersic index $n$ is measured independently from a single-component, two-dimensional S\'ersic fit to the $r$-band image using GALFIT \citep{Peng2002GALFIT, Peng2010GALFIT}. The ring size is described by the inner and outer semi-major axes of the detected annular component. We compute stellar-mass surface densities as $\mu_{\star,{\rm core}}=M_{\star,{\rm core}}/(2\pi R_{\rm e}^2)$ for the core and $\mu_{\star,{\rm ring}}=M_{\star,{\rm ring}}/[\pi(R_{\rm out}^2-R_{\rm in}^2)]$ for the ring.

Stellar masses are estimated from the colour--mass-to-light relation of \citet{Bell2003}. We use the $z$-band luminosity and the $g-z$ colour in the form
\begin{equation}
    \log_{10}(M_\star/L_z) = a_z + b_z(g-z),
\end{equation}
with $a_z=-0.171$, $b_z=0.322$, and an adopted solar absolute magnitude $M_{z,\odot}=4.50$. These relations are calibrated for the ``diet Salpeter'' initial mass function introduced by \citet{Bell2003}, which has fewer low-mass stars than a full Salpeter IMF. To place the resulting masses on the same scale as the SDSS comparison catalogues used below, we apply a uniform $-0.19$ dex correction to the Bell et al. masses. The tabulated mass uncertainties are obtained by propagating the $g$- and $z$-band magnitude errors independently. They do not include cross-band covariance or the intrinsic scatter of the colour--mass-to-light relation.

The measured core and ring quantities are listed in Tables~\ref{tab:core_properties} and \ref{tab:ring_properties}, respectively.

\begin{table*}
    \centering
    \caption{Coordinates, redshifts, and photometric and environmental properties of the cores. Columns list the catalogue name, right ascension, declination, redshift, $g$-, $r$-, and $z$-band absolute magnitudes, stellar mass, effective semi-major axis, $R_{90}$, stellar-mass surface density, single-component S\'ersic index, matched group halo mass, central--satellite flag, and ELUCID overdensity. Except for the coordinates, redshift, $n_{\rm core}$, $\log M_{\rm halo}$, central--satellite flag, and $\delta$, values are reported with their measurement uncertainties.}
    \label{tab:core_properties}
    \fontsize{5.5}{6.6}\selectfont
    \setlength{\tabcolsep}{0.4pt}
    \begin{tabular}{lcccccccccccccc}
        \hline
        Name & RA & Dec. & $z$ & $M_{g,{\rm core}}$ & $M_{r,{\rm core}}$ & $M_{z,{\rm core}}$ & $\log M_{\star,{\rm core}}$ & $R_{\rm e,core}$ & $R_{90,{\rm core}}$ & $\log \mu_{\star,{\rm core}}$ & $n_{\rm core}$ & $\log M_{\rm halo}$ & Central & $\delta$ \\
        & deg & deg & & mag & mag & mag & $M_\odot$ & kpc & kpc & $M_\odot\,{\rm kpc}^{-2}$ & & $M_\odot$ & & \\
        \hline
        SDSS J004228.79+161728.5 & $10.6200$ & $16.2913$ & $0.079$ & $-20.30\pm 0.003$ & $-21.20\pm 0.004$ & $-21.83\pm 0.004$ & $10.67\pm 0.003$ & $2.56\pm 0.013$ & $7.39\pm 0.051$ & $9.05\pm 0.005$ & $5.58$ & -- & -- & -- \\
        LEDA 1818512 & $122.9412$ & $27.8073$ & $0.076$ & $-20.59\pm 0.008$ & $-21.42\pm 0.008$ & $-22.09\pm 0.007$ & $10.76\pm 0.006$ & $3.71\pm 0.031$ & $10.42\pm 0.089$ & $8.82\pm 0.009$ & $5.89$ & $12.70$ & yes & $1.81$ \\
        SDSS J082823.59+015614.6 & $127.0983$ & $1.9374$ & $0.086$ & $-20.60\pm 0.022$ & $-21.55\pm 0.002$ & $-22.19\pm 0.003$ & $10.83\pm 0.007$ & $2.61\pm 0.008$ & $6.81\pm 0.037$ & $9.20\pm 0.008$ & $3.76$ & -- & -- & -- \\
        UGC 4599 & $131.9238$ & $13.4192$ & $0.007$ & $-18.24\pm 0.011$ & $-18.92\pm 0.008$ & $-19.40\pm 0.006$ & $9.57\pm 0.006$ & $1.32\pm 0.012$ & $4.10\pm 0.044$ & $8.53\pm 0.010$ & $3.36$ & -- & -- & -- \\
        LEDA 1668413 & $133.2369$ & $22.3571$ & $0.038$ & $-19.81\pm 0.005$ & $-20.60\pm 0.004$ & $-21.15\pm 0.005$ & $10.33\pm 0.004$ & $1.98\pm 0.009$ & $5.05\pm 0.025$ & $8.93\pm 0.006$ & $3.65$ & $12.14$ & yes & $1.02$ \\
        LEDA 3097463 & $135.0518$ & $38.8762$ & $0.075$ & $-19.39\pm 0.004$ & $-20.31\pm 0.004$ & $-20.87\pm 0.004$ & $10.26\pm 0.003$ & $2.03\pm 0.009$ & $6.23\pm 0.045$ & $8.85\pm 0.005$ & $13.94$ & $12.14$ & yes & $0.69$ \\
        LEDA 1829531 & $13.4248$ & $28.2166$ & $0.076$ & $-20.52\pm 0.007$ & $-21.37\pm 0.006$ & $-21.98\pm 0.033$ & $10.70\pm 0.024$ & $3.64\pm 0.024$ & $9.25\pm 0.104$ & $8.78\pm 0.025$ & $4.16$ & -- & -- & -- \\
        LEDA 1564718 & $150.9724$ & $18.6114$ & $0.079$ & $-21.11\pm 0.007$ & $-22.03\pm 0.003$ & $-22.66\pm 0.005$ & $11.00\pm 0.004$ & $4.11\pm 0.012$ & $11.87\pm 0.025$ & $8.97\pm 0.005$ & $3.09$ & $12.89$ & yes & $2.67$ \\
        SDSS J010054.43+253637.1 & $15.2268$ & $25.6103$ & $0.079$ & $-20.42\pm 0.006$ & $-21.34\pm 0.005$ & $-21.96\pm 0.006$ & $10.72\pm 0.005$ & $2.75\pm 0.016$ & $7.59\pm 0.065$ & $9.04\pm 0.007$ & $4.59$ & -- & -- & -- \\
        LEDA 1840343 & $163.5727$ & $28.5936$ & $0.077$ & $-19.89\pm 0.003$ & $-20.82\pm 0.003$ & $-21.45\pm 0.003$ & $10.52\pm 0.002$ & $2.11\pm 0.005$ & $5.55\pm 0.021$ & $9.07\pm 0.003$ & $2.81$ & $12.23$ & yes & $0.67$ \\
        LEDA 986443 & $164.7093$ & $-9.7864$ & $0.0678$ & $-20.55\pm 0.004$ & $-21.40\pm 0.004$ & $-22.05\pm 0.003$ & $10.74\pm 0.003$ & $2.90\pm 0.015$ & $7.56\pm 0.038$ & $9.02\pm 0.005$ & $3.18$ & -- & -- & -- \\
        PGC 1000714 & $170.8185$ & $-8.6685$ & $0.02574$ & $-19.38\pm 0.009$ & $-20.14\pm 0.009$ & $-20.69\pm 0.009$ & $10.14\pm 0.007$ & $1.95\pm 0.021$ & $6.31\pm 0.059$ & $8.76\pm 0.011$ & $4.61$ & -- & -- & -- \\
        Z 39-164 & $170.9797$ & $3.0987$ & $0.045$ & $-20.41\pm 0.002$ & $-21.25\pm 0.002$ & $-21.86\pm 0.002$ & $10.65\pm 0.001$ & $1.94\pm 0.006$ & $5.51\pm 0.024$ & $9.28\pm 0.003$ & $3.03$ & $12.69$ & yes & $0.96$ \\
        LEDA 4547778 & $174.3069$ & $23.2894$ & $0.049$ & $-18.44\pm 0.009$ & $-19.29\pm 0.031$ & $-19.87\pm 0.007$ & $9.85\pm 0.006$ & $1.26\pm 0.035$ & $3.37\pm 0.236$ & $8.85\pm 0.025$ & $3.19$ & -- & yes & $1.50$ \\
        2MASS J12014342+0010593 & $180.4311$ & $0.1831$ & $0.104$ & $-18.90\pm 0.009$ & $-19.85\pm 0.010$ & $-20.51\pm 0.035$ & $10.16\pm 0.026$ & $1.93\pm 0.013$ & $4.41\pm 0.071$ & $8.79\pm 0.026$ & $2.21$ & -- & -- & $1.21$ \\
        2MASX J12363134+2325332 & $189.1307$ & $23.4261$ & $0.046$ & $-20.07\pm 0.004$ & $-20.91\pm 0.002$ & $-21.53\pm 0.003$ & $10.52\pm 0.002$ & $2.22\pm 0.007$ & $5.59\pm 0.018$ & $9.03\pm 0.004$ & $6.53$ & $12.37$ & yes & $0.65$ \\
        Z 70-200 & $189.5933$ & $9.5283$ & $0.032$ & $-20.25\pm 0.005$ & $-21.05\pm 0.004$ & $-21.63\pm 0.003$ & $10.53\pm 0.003$ & $2.00\pm 0.011$ & $5.48\pm 0.018$ & $9.13\pm 0.006$ & $3.16$ & $12.43$ & yes & $0.44$ \\
        SDSS J125253.45+531427.1 & $193.2227$ & $53.2409$ & $0.055$ & $-18.38\pm 0.012$ & $-19.20\pm 0.011$ & $-19.72\pm 0.009$ & $9.76\pm 0.008$ & $1.40\pm 0.012$ & $3.80\pm 0.039$ & $8.67\pm 0.011$ & $7.43$ & $13.89$ & no & $2.23$ \\
        2MASX J13130376+1050115 & $198.2657$ & $10.8366$ & $0.054$ & $-20.31\pm 0.004$ & $-21.15\pm 0.003$ & $-21.77\pm 0.002$ & $10.62\pm 0.002$ & $2.13\pm 0.009$ & $5.72\pm 0.022$ & $9.17\pm 0.004$ & $2.30$ & $12.56$ & yes & $2.79$ \\
        SDSS J011843.87+250129.9 & $19.6828$ & $25.0250$ & $0.066$ & $-19.85\pm 0.014$ & $-20.73\pm 0.003$ & $-21.36\pm 0.005$ & $10.47\pm 0.006$ & $2.68\pm 0.007$ & $6.40\pm 0.025$ & $8.82\pm 0.006$ & $2.02$ & -- & -- & -- \\
        SDSSCGB 8233.1 & $200.0945$ & $-0.4021$ & $0.11$ & $-19.54\pm 0.015$ & $-20.54\pm 0.008$ & $-21.22\pm 0.009$ & $10.47\pm 0.008$ & $2.26\pm 0.012$ & $5.28\pm 0.058$ & $8.96\pm 0.009$ & $3.36$ & $12.48$ & yes & $5.38$ \\
        SDSS J132549.22-015957.5 & $201.4551$ & $-1.9993$ & $0.084$ & $-19.96\pm 0.007$ & $-20.87\pm 0.004$ & $-21.50\pm 0.004$ & $10.54\pm 0.004$ & $2.59\pm 0.010$ & $6.53\pm 0.029$ & $8.91\pm 0.005$ & $3.30$ & $12.32$ & yes & $5.94$ \\
        ZW VIII 437 & $219.0084$ & $2.8516$ & $0.029$ & $-20.31\pm 0.015$ & $-21.09\pm 0.010$ & $-21.64\pm 0.016$ & $10.52\pm 0.013$ & $2.80\pm 0.038$ & $9.19\pm 0.110$ & $8.83\pm 0.017$ & $6.59$ & $12.31$ & yes & $2.62$ \\
        WISEA J151059.53-043028.0 & $227.7480$ & $-4.5078$ & $0.084$ & $-21.10\pm 0.005$ & $-22.03\pm 0.003$ & $-22.67\pm 0.003$ & $11.02\pm 0.003$ & $3.57\pm 0.013$ & $10.55\pm 0.049$ & $9.11\pm 0.004$ & $2.64$ & -- & -- & -- \\
        Hoag's Object & $229.3083$ & $21.5856$ & $0.042$ & $-20.02\pm 0.003$ & $-20.87\pm 0.003$ & $-21.50\pm 0.002$ & $10.51\pm 0.002$ & $1.76\pm 0.005$ & $5.33\pm 0.017$ & $9.22\pm 0.003$ & $3.59$ & -- & -- & $3.13$ \\
        LEDA 1543950 & $23.8568$ & $17.8684$ & $0.068$ & $-19.75\pm 0.011$ & $-20.59\pm 0.007$ & $-21.25\pm 0.007$ & $10.42\pm 0.006$ & $2.57\pm 0.017$ & $6.94\pm 0.065$ & $8.80\pm 0.009$ & $2.65$ & -- & -- & -- \\
        NGC 6028 & $240.3704$ & $19.3600$ & $0.015$ & $-19.61\pm 0.005$ & $-20.43\pm 0.006$ & $-20.98\pm 0.004$ & $10.28\pm 0.004$ & $1.22\pm 0.020$ & $4.00\pm 0.030$ & $9.31\pm 0.014$ & $6.46$ & $12.37$ & yes & $0.73$ \\
        2MASX J01514310+0524439 & $27.9297$ & $5.4123$ & $0.061$ & $-19.60\pm 0.005$ & $-20.49\pm 0.002$ & $-21.15\pm 0.002$ & $10.40\pm 0.002$ & $2.21\pm 0.006$ & $6.58\pm 0.030$ & $8.91\pm 0.003$ & $3.34$ & -- & -- & -- \\
        LEDA 3372664 & $322.7024$ & $10.5784$ & $0.063$ & $-20.08\pm 0.007$ & $-20.93\pm 0.003$ & $-21.53\pm 0.004$ & $10.51\pm 0.004$ & $2.07\pm 0.007$ & $5.87\pm 0.022$ & $9.08\pm 0.005$ & $3.72$ & $13.90$ & no & -- \\
        SDSS J215039.21+204249.0 & $327.6634$ & $20.7136$ & $0.074$ & $-20.50\pm 0.008$ & $-21.40\pm 0.005$ & $-22.07\pm 0.006$ & $10.77\pm 0.005$ & $3.15\pm 0.019$ & $8.17\pm 0.052$ & $8.97\pm 0.007$ & $3.30$ & -- & -- & -- \\
        \hline
    \end{tabular}
\end{table*}

\begin{table*}
    \centering
    \caption{Photometric and structural properties of the rings. Columns list the catalogue name, $g$-, $r$-, and $z$-band absolute magnitudes, stellar mass, inner and outer semi-major axes, and stellar-mass surface density. All values are reported with their measurement uncertainties.}
    \label{tab:ring_properties}
    \fontsize{5.5}{6.6}\selectfont
    \setlength{\tabcolsep}{1.8pt}
    \begin{tabular}{lccccccc}
        \hline
        Name & $M_{g,{\rm ring}}$ & $M_{r,{\rm ring}}$ & $M_{z,{\rm ring}}$ & $\log M_{\star,{\rm ring}}$ & $R_{\rm in}$ & $R_{\rm out}$ & $\log \mu_{\star,{\rm ring}}$ \\
        & mag & mag & mag & $M_\odot$ & kpc & kpc & $M_\odot\,{\rm kpc}^{-2}$ \\
        \hline
        SDSS J004228.79+161728.5 & $-19.41\pm 0.016$ & $-20.14\pm 0.029$ & $-20.67\pm 0.031$ & $10.12\pm 0.023$ & $6.15\pm 0.054$ & $19.25\pm 0.337$ & $7.10\pm 0.028$ \\
        LEDA 1818512 & $-18.56\pm 0.047$ & $-19.04\pm 0.080$ & $-19.56\pm 0.077$ & $9.58\pm 0.057$ & $10.21\pm 0.182$ & $19.94\pm 0.375$ & $6.62\pm 0.062$ \\
        SDSS J082823.59+015614.6 & $-19.96\pm 0.060$ & $-20.80\pm 0.018$ & $-21.32\pm 0.029$ & $10.40\pm 0.029$ & $6.34\pm 0.103$ & $25.72\pm 0.946$ & $7.11\pm 0.044$ \\
        UGC 4599 & $-16.89\pm 0.183$ & $-17.38\pm 0.092$ & $-17.67\pm 0.087$ & $8.76\pm 0.086$ & $5.00\pm 0.184$ & $12.73\pm 0.396$ & $6.12\pm 0.092$ \\
        LEDA 1668413 & $-18.30\pm 0.061$ & $-19.06\pm 0.037$ & $-19.44\pm 0.080$ & $9.58\pm 0.061$ & $6.12\pm 0.097$ & $17.54\pm 0.175$ & $6.66\pm 0.062$ \\
        LEDA 3097463 & $-19.62\pm 0.023$ & $-20.07\pm 0.028$ & $-20.23\pm 0.035$ & $9.73\pm 0.026$ & $6.09\pm 0.174$ & $26.27\pm 0.388$ & $6.41\pm 0.029$ \\
        LEDA 1829531 & $-19.44\pm 0.030$ & $-20.13\pm 0.018$ & $-20.58\pm 0.092$ & $10.04\pm 0.067$ & $11.59\pm 0.163$ & $24.21\pm 0.479$ & $6.89\pm 0.071$ \\
        LEDA 1564718 & $-19.25\pm 0.075$ & $-19.94\pm 0.059$ & $-20.24\pm 0.114$ & $9.85\pm 0.086$ & $9.16\pm 0.165$ & $25.25\pm 0.402$ & $6.61\pm 0.087$ \\
        SDSS J010054.43+253637.1 & $-19.63\pm 0.025$ & $-20.40\pm 0.028$ & $-21.02\pm 0.022$ & $10.29\pm 0.018$ & $6.55\pm 0.060$ & $20.63\pm 1.527$ & $7.21\pm 0.074$ \\
        LEDA 1840343 & $-19.18\pm 0.038$ & $-19.67\pm 0.051$ & $-20.14\pm 0.069$ & $9.80\pm 0.051$ & $11.26\pm 0.161$ & $27.57\pm 0.655$ & $6.50\pm 0.057$ \\
        LEDA 986443 & $-19.33\pm 0.023$ & $-20.11\pm 0.039$ & $-20.74\pm 0.021$ & $10.19\pm 0.017$ & $8.98\pm 0.108$ & $24.42\pm 0.337$ & $6.98\pm 0.022$ \\
        PGC 1000714 & $-17.80\pm 0.111$ & $-18.09\pm 0.160$ & $-18.59\pm 0.180$ & $9.13\pm 0.135$ & $10.49\pm 0.413$ & $19.53\pm 0.543$ & $6.20\pm 0.140$ \\
        Z 39-164 & $-19.98\pm 0.012$ & $-20.69\pm 0.016$ & $-21.19\pm 0.010$ & $10.31\pm 0.008$ & $4.15\pm 0.135$ & $22.80\pm 0.243$ & $7.11\pm 0.013$ \\
        LEDA 4547778 & $-17.87\pm 0.084$ & $-18.55\pm 0.072$ & $-18.86\pm 0.096$ & $9.30\pm 0.075$ & $6.81\pm 0.169$ & $15.06\pm 0.459$ & $6.55\pm 0.082$ \\
        2MASS J12014342+0010593 & $-19.94\pm 0.023$ & $-20.39\pm 0.040$ & $-20.74\pm 0.100$ & $9.99\pm 0.073$ & $7.00\pm 0.125$ & $20.71\pm 0.222$ & $6.91\pm 0.073$ \\
        2MASX J12363134+2325332 & $-19.69\pm 0.020$ & $-20.42\pm 0.015$ & $-20.94\pm 0.018$ & $10.22\pm 0.015$ & $4.14\pm 0.077$ & $18.22\pm 0.235$ & $7.22\pm 0.019$ \\
        Z 70-200 & $-18.59\pm 0.086$ & $-19.37\pm 0.068$ & $-19.87\pm 0.065$ & $9.80\pm 0.055$ & $6.60\pm 0.070$ & $17.74\pm 0.165$ & $6.87\pm 0.056$ \\
        SDSS J125253.45+531427.1 & $-18.61\pm 0.202$ & $-19.02\pm 0.074$ & $-19.27\pm 0.071$ & $9.36\pm 0.083$ & $4.14\pm 0.069$ & $14.09\pm 0.611$ & $6.60\pm 0.093$ \\
        2MASX J13130376+1050115 & $-19.59\pm 0.027$ & $-20.25\pm 0.033$ & $-20.85\pm 0.019$ & $10.19\pm 0.016$ & $5.35\pm 0.049$ & $17.05\pm 0.579$ & $7.27\pm 0.036$ \\
        SDSS J011843.87+250129.9 & $-19.42\pm 0.065$ & $-20.02\pm 0.022$ & $-20.49\pm 0.050$ & $9.98\pm 0.042$ & $6.92\pm 0.059$ & $19.87\pm 0.432$ & $6.94\pm 0.047$ \\
        SDSSCGB 8233.1 & $-20.15\pm 0.081$ & $-20.60\pm 0.071$ & $-21.12\pm 0.072$ & $10.20\pm 0.058$ & $8.79\pm 0.212$ & $26.16\pm 0.915$ & $6.92\pm 0.067$ \\
        SDSS J132549.22-015957.5 & $-19.26\pm 0.087$ & $-19.79\pm 0.069$ & $-20.22\pm 0.082$ & $9.84\pm 0.065$ & $7.96\pm 0.728$ & $24.69\pm 0.850$ & $6.60\pm 0.074$ \\
        ZW VIII 437 & $-18.11\pm 0.095$ & $-18.98\pm 0.042$ & $-19.73\pm 0.063$ & $9.85\pm 0.055$ & $4.89\pm 0.077$ & $12.31\pm 0.204$ & $7.25\pm 0.058$ \\
        WISEA J151059.53-043028.0 & $-19.17\pm 0.067$ & $-19.79\pm 0.054$ & $-20.09\pm 0.085$ & $9.77\pm 0.065$ & $12.63\pm 0.144$ & $26.46\pm 1.272$ & $6.54\pm 0.085$ \\
        Hoag's Object & $-20.13\pm 0.022$ & $-20.64\pm 0.025$ & $-20.94\pm 0.031$ & $10.08\pm 0.024$ & $9.71\pm 0.273$ & $24.98\pm 0.303$ & $6.86\pm 0.027$ \\
        LEDA 1543950 & $-19.40\pm 0.046$ & $-20.04\pm 0.034$ & $-20.56\pm 0.047$ & $10.04\pm 0.037$ & $6.57\pm 0.056$ & $20.16\pm 0.556$ & $6.98\pm 0.046$ \\
        NGC 6028 & $-18.98\pm 0.026$ & $-19.49\pm 0.036$ & $-19.92\pm 0.035$ & $9.71\pm 0.027$ & $4.78\pm 0.084$ & $14.54\pm 0.389$ & $6.93\pm 0.037$ \\
        2MASX J01514310+0524439 & $-19.96\pm 0.020$ & $-20.55\pm 0.009$ & $-20.98\pm 0.012$ & $10.16\pm 0.010$ & $6.87\pm 0.069$ & $27.39\pm 0.339$ & $6.82\pm 0.016$ \\
        LEDA 3372664 & $-18.43\pm 0.030$ & $-19.19\pm 0.040$ & $-19.71\pm 0.074$ & $9.74\pm 0.054$ & $7.36\pm 0.050$ & $17.39\pm 0.302$ & $6.84\pm 0.057$ \\
        SDSS J215039.21+204249.0 & $-19.85\pm 0.057$ & $-20.66\pm 0.043$ & $-21.26\pm 0.049$ & $10.39\pm 0.040$ & $9.19\pm 0.069$ & $28.45\pm 1.195$ & $7.04\pm 0.057$ \\
        \hline
    \end{tabular}
\end{table*}

\section{Scaling relations and environment}
\label{sec:results}

In this section, we place the photometric properties of Hoag-type galaxies in the context of normal low-redshift galaxies. We first construct the SDSS samples used for comparison (Section~\ref{subsec:sdss_comparison}), and then examine the locations of the cores and rings in the colour--magnitude diagram (Section~\ref{subsec:photometric_distributions}). We next compare the colours and structural scaling relations of the cores with those of normal galaxies (Section~\ref{subsec:core_scaling}). Finally, we investigate the group membership and large-scale environments of the Hoag-type systems (Section~\ref{subsec:group_environment}).

\subsection{SDSS comparison sample}
\label{subsec:sdss_comparison}

We construct an SDSS comparison sample in order to place the Hoag-type cores and rings in the context of normal low-redshift galaxies. The parent catalogue is drawn from the New York University Value-Added Galaxy Catalog\footnote{\url{https://sdss.physics.nyu.edu/vagc/}} \citep[NYU-VAGC;][]{Blanton2005}, which provides value-added reductions and derived quantities for galaxies in public survey data. We use the low-redshift galaxy sample in the NYU-VAGC, which contains of order half a million galaxies with spectroscopic redshifts and $r$-band Petrosian apparent magnitudes, selected with $r<17.6$ and $0.01<z<0.2$. From this parent sample we construct a volume-limited comparison sample by requiring $\log_{10}(M_\star/M_\odot)>9$, $0.015<z<0.035$, yielding 31215 galaxies. We refer to this volume-limited SDSS sample as SampleA. Optical photometric quantities are taken from the NASA-Sloan Atlas\footnote{\url{https://www.nsatlas.org/}} \citep[NSA;][]{Blanton2011}.

For structural comparisons, we use the NSA elliptical Petrosian quantities, specifically \texttt{ELPETRO\_TH50} and \texttt{ELPETRO\_TH90}, for the comparison-galaxy half-light and 90-per-cent-light radii \citep{Blanton2011,Wake2017}. The concentration of the comparison galaxies is then defined as \texttt{ELPETRO\_TH90}/\texttt{ELPETRO\_TH50}. We also include the NSA $r$-band S\'ersic index, measured from two-dimensional single-component S\'ersic fits, as an additional structural parameter from the same rephotometry \citep{Blanton2011}.

We further assign morphological information to the SDSS comparison galaxies by matching to the SDSS structural catalogue of \citet{Meert2015} and to the deep-learning morphology catalogue of \citet{DominguezSanchez2018}, which provides Hubble-sequence T-type estimates for SDSS galaxies. We divide SampleA into an early-type subset with $T\leq0$, referred to as SampleE, and a late-type subset with $T>0$, referred to as SampleL. The visual definition used for our Hoag-type sample requires a smooth central component without obvious spiral arms or a strong bar, so the core component is naturally compared with early-type galaxies rather than with the full star-forming disc population. Since stellar mass is the primary variable controlling many galaxy scaling relations, we construct an additional mass-matched early-type control sample from SampleE. For each Hoag-type core, we randomly select ten galaxies with $|\Delta \log_{10} M_\star|<0.11$ dex relative to the core stellar mass. This procedure gives a control sample of 300 early-type galaxy entries, hereafter SampleC, which is used to compare the cores of Hoag-type galaxies with normal galaxies of similar stellar mass and morphology.

\subsection{Colour--magnitude diagram}
\label{subsec:photometric_distributions}

\begin{figure}
    \includegraphics[width=\columnwidth]{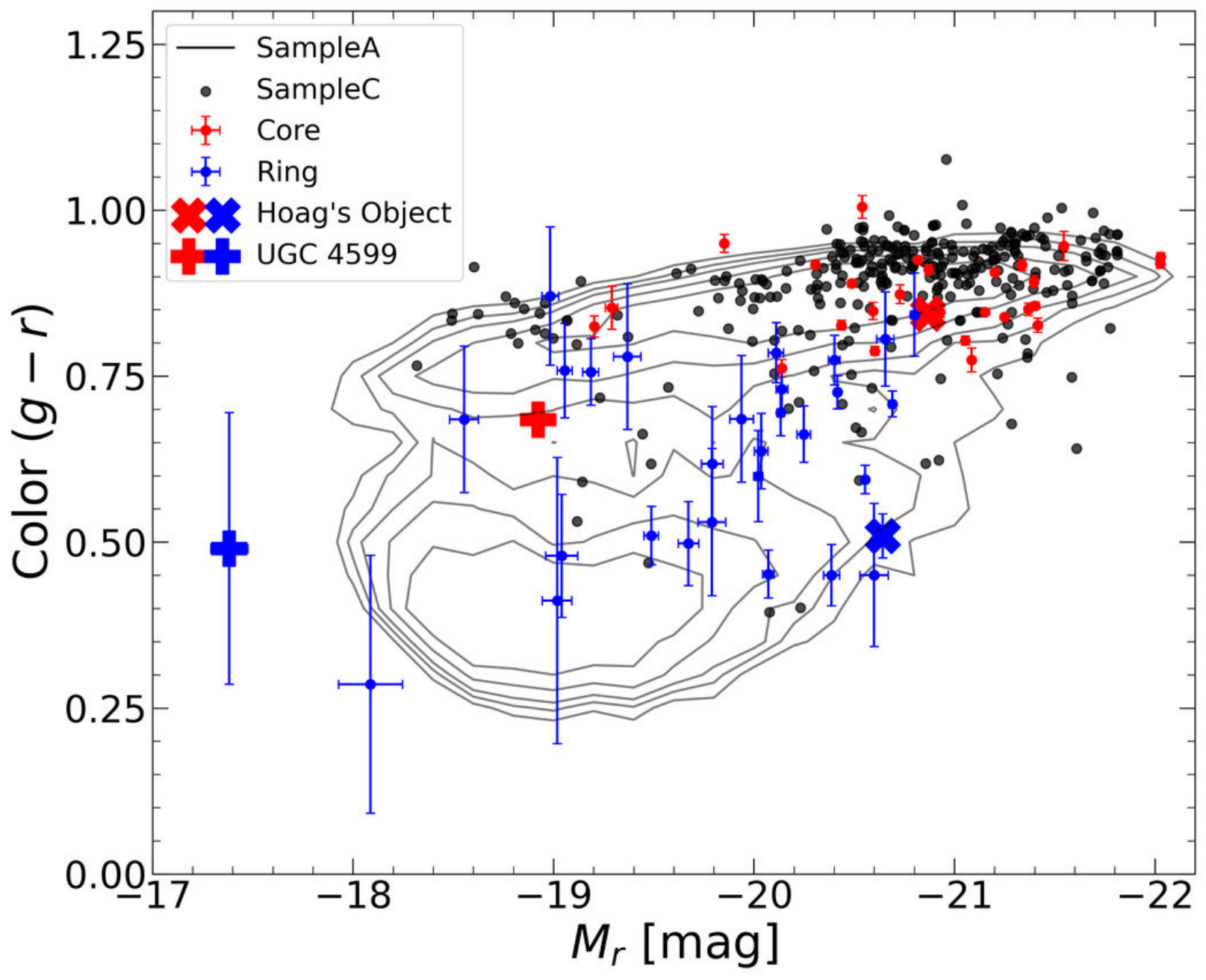}
    \caption{Colour--magnitude diagram for the Hoag-type components and the SDSS comparison samples. Grey contours show SampleA, black points show SampleC, red points show Hoag-type cores, and blue points show Hoag-type rings. Error bars indicate measurement uncertainties for the Hoag-type components. The large crosses and plus signs mark Hoag's Object and UGC 4599, respectively, with red and blue indicating their cores and rings.}
    \label{fig:mag_colour}
\end{figure}

Figure~\ref{fig:mag_colour} shows the $g-r$ colour versus $r$-band absolute magnitude diagram for the Hoag-type components, SampleA, and SampleC. In low-redshift galaxy samples, colour--magnitude diagrams show a well-known bimodality: a red sequence dominated by quiescent, early-type systems and a blue cloud dominated by star-forming disc galaxies \citep{Strateva2001,Baldry2004}. This plane therefore provides a simple first view of the stellar populations of the decomposed cores and rings.

The grey contours in Figure~\ref{fig:mag_colour} show SampleA, while the black points mark SampleC. The Hoag-type cores, shown as red points, lie mainly on the red sequence and overlap the locus of SampleC. This supports the interpretation that the central components are dominated by relatively old and/or quenched stellar populations, consistent with the smooth early-type morphology imposed by the visual selection. Three objects---UGC 4599, LEDA 4547778, and SDSS J125253.45+531427.1---stand out at the low-luminosity end of the core distribution. We refer to these objects as the three core outliers in the subsequent scatter plots and examine their influence on the statistical comparisons in Section~\ref{subsec:sample_purity}. By contrast, the rings, shown as blue points, occupy a much broader range in colour. Some rings lie close to the red sequence, but many extend toward the blue cloud. This spread indicates that the outer rings are not a uniformly old stellar component: at least part of the ring population contains younger stellar populations or more active star formation than the cores. The coexistence of red-sequence cores with rings spanning from red to blue colours suggests that the cores and rings have different recent star-formation histories, even when they are observed in the same galaxies.

\subsection{Core scaling relations}
\label{subsec:core_scaling}

\begin{figure*}
    \includegraphics[width=\textwidth]{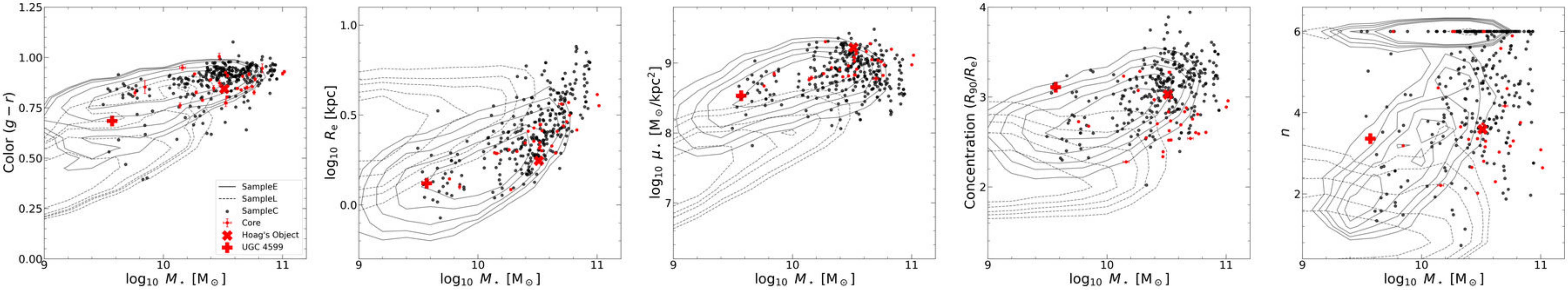}
    \caption{Core stellar mass compared with core colour, half-light radius, stellar mass surface density, concentration, and S\'ersic index. Solid grey contours show SampleE, dashed grey contours show SampleL, black points show SampleC, and red points with error bars show Hoag-type cores. The large red crosses and plus signs mark the cores of Hoag's Object and UGC 4599, respectively.}
    \label{fig:core_mass_params}
\end{figure*}

\begin{figure*}
    \includegraphics[width=\textwidth]{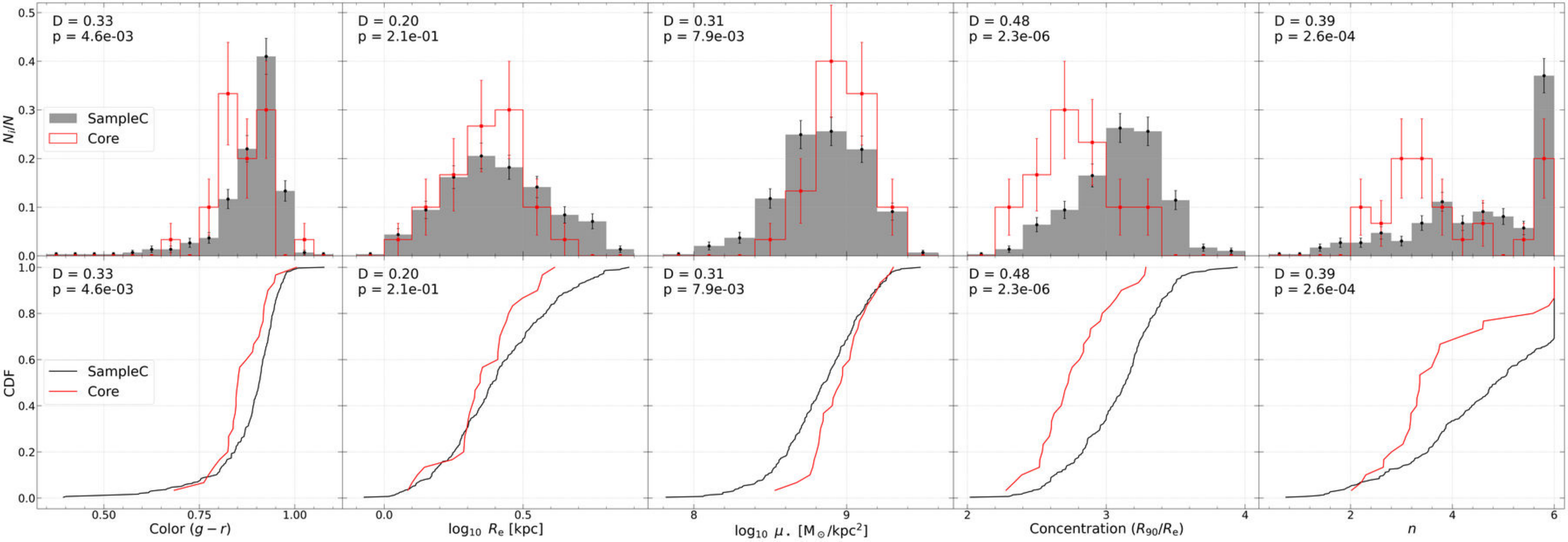}
    \caption{Distributions of core colour, half-light radius, stellar mass surface density, concentration, and S\'ersic index for Hoag-type cores and SampleC. The upper panels show normalized histograms and the lower panels show cumulative distribution functions. The reported $D$ and $p$ values are from two-sample Kolmogorov--Smirnov tests.}
    \label{fig:core_params_1}
\end{figure*}

Because stellar mass is the primary scale on which many galaxy properties vary, we first examine whether the structural and stellar-population properties of the Hoag-type cores follow the same mass-dependent trends as normal low-redshift galaxies. Figure~\ref{fig:core_mass_params} compares core colour, size, stellar mass surface density, concentration, and S\'ersic index with stellar mass. The three core outliers identified in Figure~\ref{fig:mag_colour} occupy the low-mass end of these distributions (Table~\ref{tab:core_properties}). For consistency with the upper limit of $n=6$ in the NSA comparison data, we cap the GALFIT-derived core Sérsic indices at $n=6$ when constructing the comparison plots and performing the Kolmogorov–Smirnov tests. Table~\ref{tab:core_properties} retains the original fitted values. The solid contours show SampleE and the dashed contours show SampleL. The black points show SampleC, while the red points show the Hoag-type cores. To make the comparison with early-type galaxies of similar mass more explicit, Figure~\ref{fig:core_params_1} shows the corresponding normalized histograms and cumulative distribution functions for the Hoag-type cores and SampleC, together with two-sample Kolmogorov--Smirnov statistics.

The cores broadly follow the mass-dependent trends of early-type galaxies: their colours lie near the red sequence, and their sizes and stellar-mass surface densities occupy the same broad regions as SampleE and SampleC, consistent with the general early-type size--mass and surface-density trends in large SDSS samples \citep{Shen2003,Kauffmann2003Structure}. However, the mass-matched distributions in Figure~\ref{fig:core_params_1} reveal offsets within these broad relations. Compared with SampleC, the cores tend to be slightly bluer and smaller, have higher stellar-mass surface densities, and have lower concentrations and S\'ersic indices. The smaller sizes and higher surface densities are related through $\mu_\star=M_\star/(2\pi R_{\rm e}^2)$ and therefore do not constitute independent structural differences. The strengths of the distribution differences vary among parameters, as indicated by the Kolmogorov--Smirnov statistics in the figure. In particular, the size difference is not statistically significant ($D=0.20$, $p=0.21$). Although the two samples are not explicitly matched in redshift, both lie in the nearby Universe, where differential evolutionary effects are expected to be small.

\begin{figure}
    \includegraphics[width=\columnwidth]{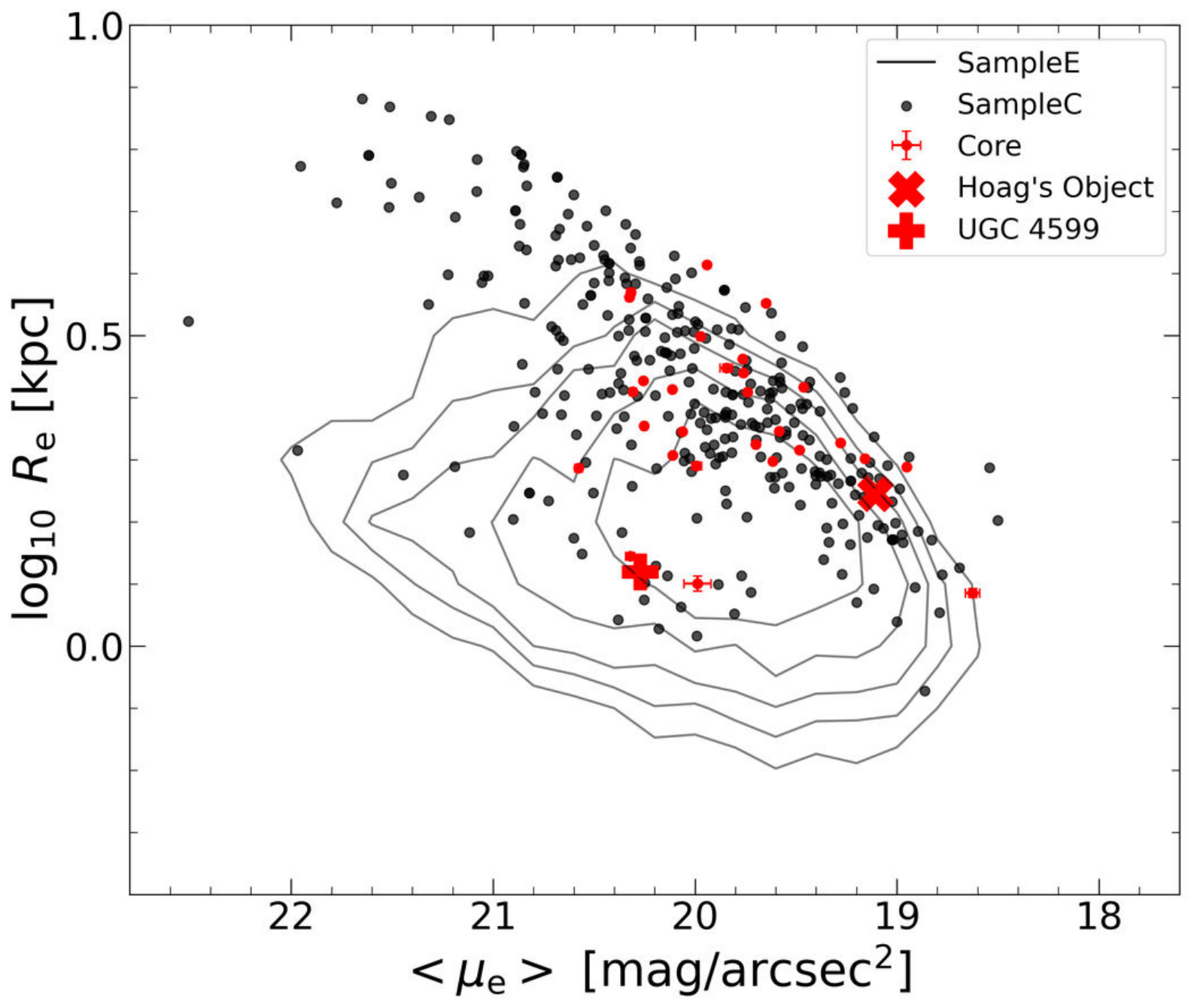}
    \caption{Effective radius versus mean $r$-band surface brightness within the effective radius. Grey contours show SampleE, black points show SampleC, and red points with error bars show Hoag-type cores. The large red crosses and plus signs mark the cores of Hoag's Object and UGC 4599, respectively.}
    \label{fig:core_sb_radius}
\end{figure}

The slightly bluer colours and lower concentrations and S\'ersic indices of the Hoag-type cores motivate a comparison with pseudobulges: \citet{Gadotti2009} found bluer colours for pseudobulges than for classical bulges, while their disc-like, near-exponential light profiles \citep{KormendyKennicutt2004} and typically low S\'ersic indices \citep{FisherDrory2008} are consistent with less centrally concentrated light distributions. However, the higher stellar-mass surface densities of the cores at fixed stellar mass suggest that they may be denser systems than the early-type comparison galaxies. We therefore examine the Kormendy diagram in Figure~\ref{fig:core_sb_radius}, comparing effective radius with the mean $r$-band surface brightness within it, $\langle\mu_{\rm e}\rangle$. The cores tend to have higher mean surface brightness at fixed $R_{\rm e}$. This is opposite to the low-surface-brightness displacement at fixed radius used by \citet{Gadotti2009} to identify pseudobulges as outliers from the elliptical/classical-bulge Kormendy relation. Although all fitted core S\'ersic indices exceed $n=2$ (Table~\ref{tab:core_properties}), this parameter alone is insufficient to classify their bulge type. Their high surface densities and surface brightnesses distinguish them from the diffuse pseudobulge population. These differences suggest that the cores may have experienced a different history of structural assembly from ordinary early-type galaxies, despite broadly following similar mass scaling relations. However, the present broad-band photometry lacks the resolved stellar-population and kinematic information needed to establish their bulge type and evolutionary history.

\subsection{Group environment}
\label{subsec:group_environment}

We next examine whether the Hoag-type cores occupy environments similar to those of normal early-type galaxies at comparable stellar mass. For this purpose, both the Hoag-type sample and SampleA are matched to the SDSS DR7 halo-based group catalogue\footnote{\url{https://gax.sjtu.edu.cn/data/Group.html}}. The catalogue is based on the group-finding method of \citet{Yang2007}, in which galaxies are assigned to dark-matter haloes and group halo masses are estimated from the characteristic luminosity or stellar mass of each group. In the catalogue, the central galaxy is identified as the most massive member of a group. For stellar-mass matching to the comparison samples, we use the mass of the core, excluding the ring, while the central--satellite flag describes the galaxy as a group member. The matched halo masses and central--satellite flags for the Hoag-type cores are listed in Table~\ref{tab:core_properties}.

In addition to group halo mass, we use the ELUCID constrained simulation to characterize the large-scale density environment of the Hoag-type systems and SampleA \citep{Wang2016ELUCID}. ELUCID reconstructs the density field in the nearby SDSS volume with a high-resolution constrained $N$-body simulation. We adopt the environmental overdensity $\delta=\rho/\bar{\rho}-1$ from the Gaussian-smoothed reconstructed density field, using the matched reconstructed positions of galaxies in redshift space. The available overdensity measurements for the Hoag-type cores are included in Table~\ref{tab:core_properties}.

\begin{figure}
    \includegraphics[width=\columnwidth]{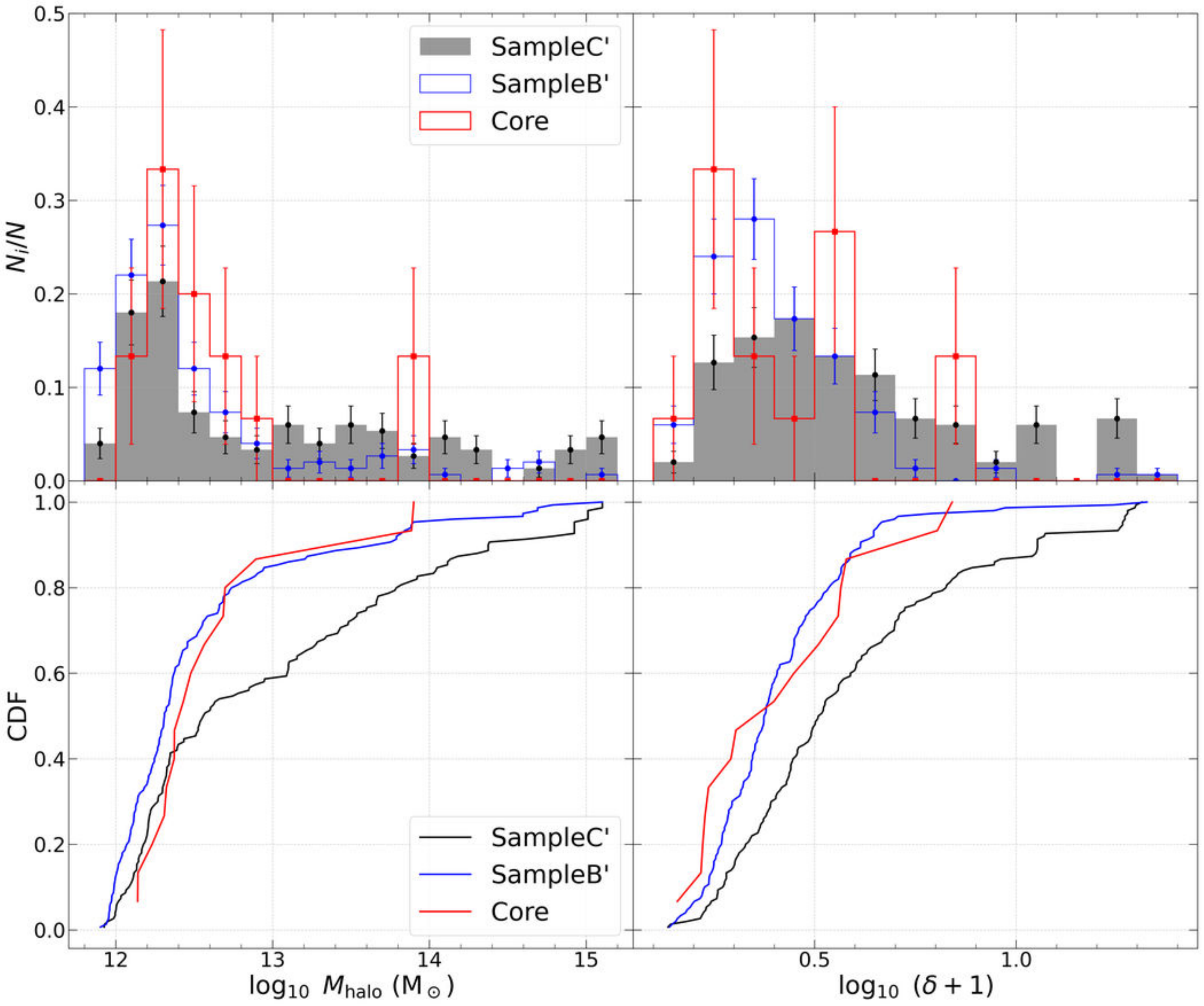}
    \caption{Distributions of group halo mass and ELUCID overdensity for Hoag-type cores, SampleC$^{\prime}$, and SampleB$^{\prime}$. SampleB$^{\prime}$ additionally matches the central–satellite flag. The upper panels show normalized histograms and the lower panels show cumulative distribution functions.}
    \label{fig:environment_distributions}
\end{figure}

Group information is available for 16 of the 30 Hoag-type galaxies. Among these matched systems, 14 are classified as centrals, 15 have valid halo-mass measurements, and 15 have valid overdensity measurements. The central fraction is 87.5 per cent, compared with 63.75 per cent (102/160) in SampleC. For the environmental comparisons, we construct two sets of control samples separately for halo mass and overdensity. The first set of controls, referred to here as the environmental SampleC$^{\prime}$, is newly drawn from SampleE using the mass-matching procedure, rather than selected as a subset of the 300 entries in Section~\ref{subsec:sdss_comparison}; only galaxies with available central–satellite flags and the relevant environmental measurements are eligible. These controls form the grey distributions in Figure~\ref{fig:environment_distributions}. Because both halo mass and overdensity depend strongly on whether a galaxy is a central or a satellite, we independently select a second set of controls with the additional requirement that their central--satellite flag matches that of the corresponding Hoag-type system. These controls, referred to as SampleB$^{\prime}$, form the blue distributions in Figure~\ref{fig:environment_distributions}.

Figure~\ref{fig:environment_distributions} compares the halo-mass and overdensity distributions of the Hoag-type cores, SampleC$^{\prime}$, and SampleB$^{\prime}$. SampleC$^{\prime}$ has a lower central fraction than the Hoag-type sample, so it contains a larger satellite contribution. In the SDSS group catalogue, halo mass is assigned to the host group rather than independently to each member galaxy \citep{Yang2007}; satellites therefore carry the halo mass of the larger system in which they reside, while centrals are associated with the halo in which they are the dominant member \citep{Yang2008}. This difference in central--satellite composition naturally shifts SampleC$^{\prime}$ toward higher assigned halo masses, and is also expected to increase the contribution from denser environments. After matching the central--satellite flag, SampleB$^{\prime}$ shows halo-mass and overdensity distributions much closer to those of the Hoag-type cores. The apparent environmental offset between the Hoag-type cores and SampleC$^{\prime}$ is therefore mainly associated with the different central--satellite flag.

The higher central fraction of the Hoag-type sample is qualitatively consistent with a reduced role for satellite-specific environmental processing. Environmental quenching preferentially affects satellite galaxies after infall into a more massive halo \citep{vanDenBosch2008,Peng2012,Wetzel2013}. We speculate that the predominance of centrals among Hoag-type systems may also favour the survival of their low-density outer rings by reducing their exposure to satellite-specific environmental processes. This is not a direct demonstration that environment formed or preserved the rings, however.

As an additional check, we cross-match our Hoag-type sample with two DESI-based group catalogues \citep{Yang2021, Williams2026}. Together, these catalogues provide 13 counterparts, including four systems already matched to the SDSS group catalogue. The nine additional systems are all classified as centrals. Combining the catalogues gives group information for 25 systems. Retaining the SDSS classifications for overlapping objects gives 23 centrals (92 per cent). SDSS J125253.45+531427.1 is classified as a satellite in SDSS but as a central in DESI group catalogue of \citet{Yang2021}\footnote{\url{https://data.desi.lbl.gov/doc/releases/dr1/vac/gfinder/}}, and SDSS J132549.22$-$015957.5 is classified as a central in SDSS but as a satellite in DESI group catalogue of \citet{Williams2026}\footnote{\url{https://data.desi.lbl.gov/doc/releases/dr1/vac/bgs-groups/}}. Therefore, adopting either of the two DESI classifications gives slightly different but similar central fractions, which are still higher than that in SampleC. Both choices remain consistent with a predominance of centrals in our Hoag-type sample. We therefore retain the SDSS-based environmental analysis above and present the DESI-based comparison as a supplementary check.

\section{Ring structure and the connection with cores}
\label{sec:ring_structure}

We now place the cores and rings within a common framework by relating the observed ring structure to the relative contributions of the core, the inner dark matter halo, and the ring itself. We define
\begin{equation}
    \eta = \frac{M_{\rm core}+M_{\rm DM}(<R_{\rm mc})}{M_{\rm ring}},
\end{equation}
where $R_{\rm mc}$ is the ring mass-centroid radius. The enclosed dark matter mass $M_{\rm DM}$ is estimated with an NFW halo profile \citep{Navarro1996}: we infer the halo mass from the stellar-to-halo mass relation of \citet{Guo2010}, assign the halo concentration using the halo mass--concentration relation of \citet{DuttonMaccio2014}, and then integrate the NFW profile within $R_{\rm mc}$. This correction is approximate, but it accounts for the fact that the orbital scale of the ring is not set by the stellar core alone. 

\begin{figure}
    \includegraphics[width=\columnwidth]{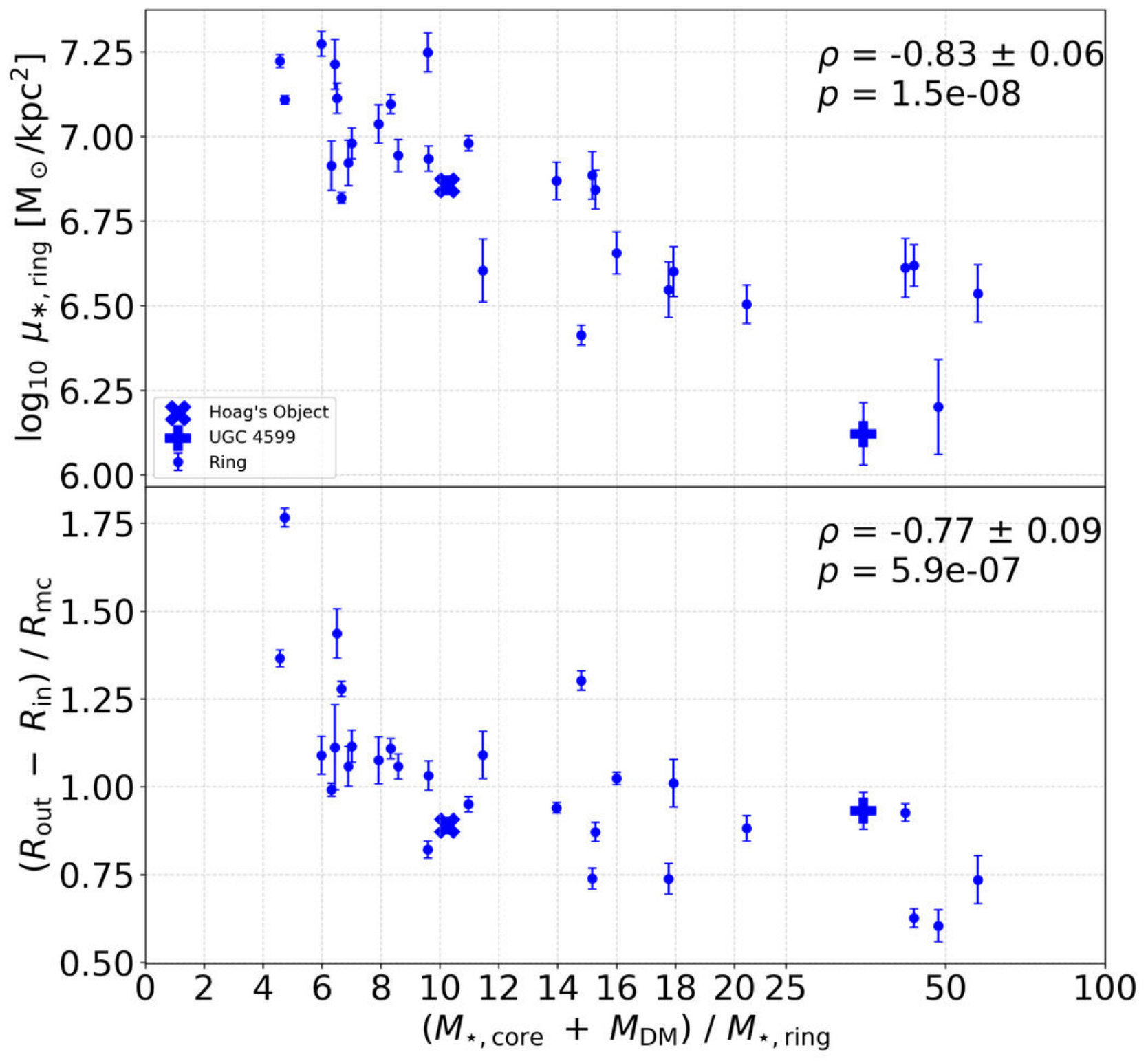}
    \caption{Ring stellar mass surface density and normalized ring width as functions of $\eta$. Error bars indicate measurement uncertainties for the Hoag-type rings. The large blue crosses and plus signs mark the rings of Hoag's Object and UGC 4599, respectively. The reported $\rho$ and $p$ values are Spearman rank statistics.}
    \label{fig:mass_ratio_ring_params}
\end{figure}

Figure~\ref{fig:mass_ratio_ring_params} shows how $\eta$ relates to the measured ring properties. The sample spans a broad range in $\eta$, from a few to about 60. The stellar mass surface density and normalized ring width $(R_{\rm out}-R_{\rm in})/R_{\rm mc}$ decrease with increasing $\eta$, with the Spearman rank coefficients shown in each panel. The measured ring widths refer to the detected annular component at the adopted signal-to-noise threshold; fainter extensions below this threshold may not be included. This suggests that systems with a larger core-plus-inner-halo mass relative to the ring tend to host lower-density and narrower rings. One possible interpretation is that $\eta$ traces different stages of ring prominence or survival, with high-$\eta$ systems corresponding to rings that are more dynamically depleted. This interpretation remains tentative, because distinguishing between formation history, subsequent fading, and dynamical evolution will require stellar-population constraints, gas measurements, and kinematic information.

\begin{figure}
    \includegraphics[width=\columnwidth]{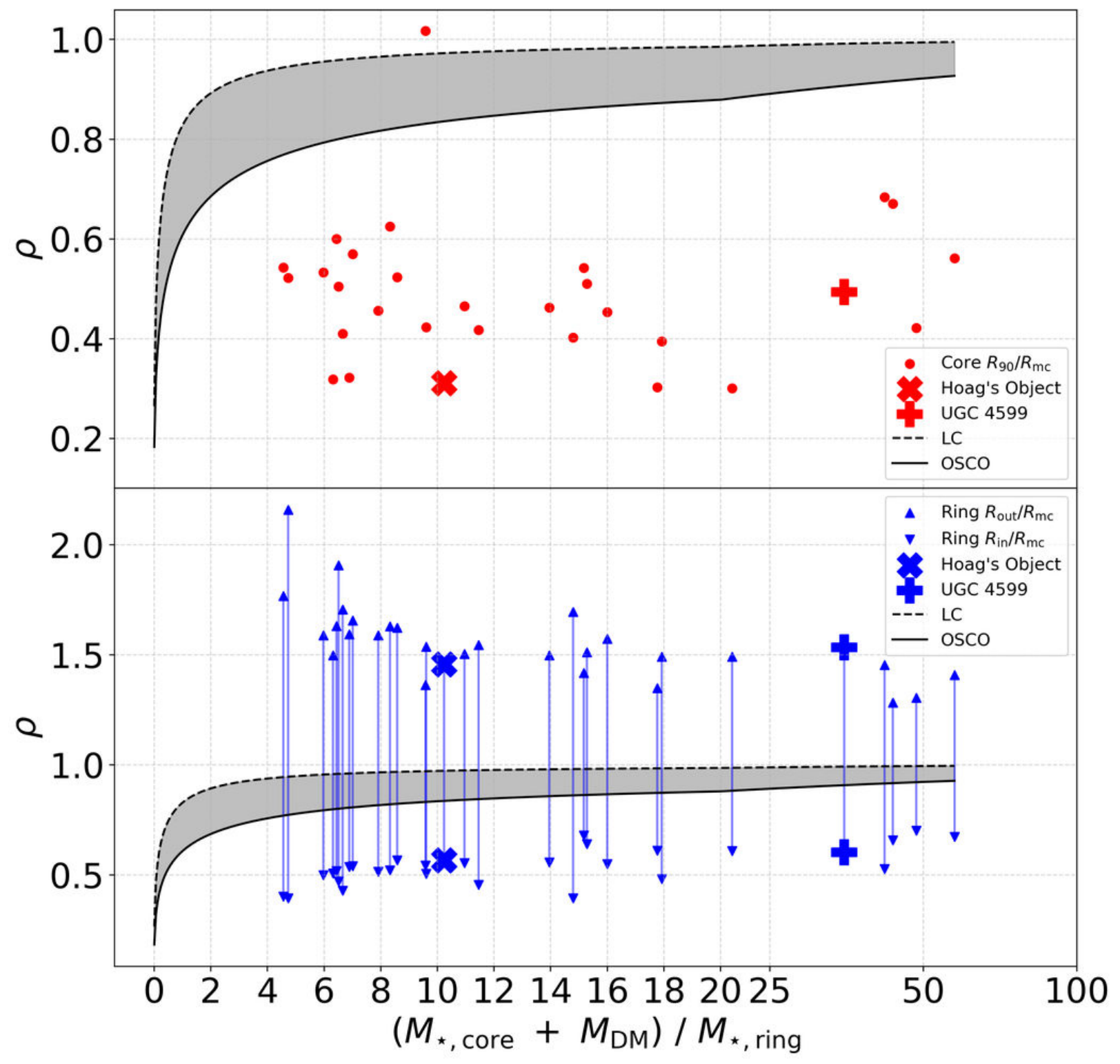}
    \caption{Comparison between observed component radii and the simplified orbital-stability model of \citet{Bannikova2018}, following the form of their Figure 8. The upper panel shows core $R_{90}/R_{\rm mc}$, and the lower panel shows the inner and outer ring radii normalized by $R_{\rm mc}$. The solid and dashed curves mark the OSCO and LC radii, respectively; the shaded region indicates the interval where stable circular orbits are absent in the simplified model. The curves adopt the effective central point mass $M_{\rm core}+M_{\rm DM}(<R_{\rm mc})$. Large crosses and plus signs mark Hoag's Object and UGC 4599, respectively.}
    \label{fig:mass_ratio_radii}
\end{figure}

We next compare these observations with the simplified dynamical model of \citet{Bannikova2018}, in which the central body is represented by a point mass and the ring by a massive circular ring. The competition between the two gravitational components defines the outermost stable circular orbit (OSCO) and the Lagrangian circle (LC); circular orbits are unstable in the region between these radii. For the comparison below, we approximate the core and the dark matter enclosed within $R_{\rm mc}$ by an effective central point mass $M_{\rm core}+M_{\rm DM}(<R_{\rm mc})$, while retaining the gravitational contribution of the massive ring. Following the logic of Figure 8 in \citet{Bannikova2018}, Figure~\ref{fig:mass_ratio_radii} compares the observed radii with the OSCO and LC curves after all radii are normalized by $R_{\rm mc}$. In the upper panel, all core $R_{90}$ values except ZW VIII 437 lie inside the OSCO curve, consistent with the central stellar component occupying the inner stable region of the simplified model. In the lower panel, the observed rings often span a broad radial range and frequently extend across the OSCO--LC interval. Thus, although the Bannikova model provides a useful reference scale for the gap between the core and ring, the observed ring locations cannot be interpreted as a direct one-to-one realization of this effective point-mass-plus-ring approximation. Additional mass components, gas dynamics, non-circular motions, or formation history may be important for setting the final ring structure.

To assess the effect of the effective point-mass approximation, we also consider a model with the radius-dependent NFW enclosed mass $M_{\rm DM}(<r)$ and the exact thin-ring force. We apply this model to the thirty objects using the tabulated component properties and approximate $R_{\rm mc}$ by the mass-weighted mean radius of an annulus with uniform surface density. Relative to the effective point-mass calculation, the median stability-radius shifts are $+0.0005R_{\rm mc}$ for the LC and $+0.0383R_{\rm mc}$ for the OSCO, with OSCO shifts ranging from $+0.0207R_{\rm mc}$ to $+0.0568R_{\rm mc}$. Most cores remain inside the OSCO, while the rings often extend across the OSCO--LC interval. These results support the use of the effective point-mass approximation in the stability analysis presented above.

\section{Discussion}
\label{sec:discussion}

\subsection{Comparison with previous studies}
\label{subsec:previous_comparison}

For the previously studied systems--Hoag's Object, NGC 6028, UGC 4599, and PGC 1000714--our measurements broadly recover the previously reported contrast between red central components and generally bluer outer structures \citep{Schweizer1987,Wakamatsu1990,Finkelman2011UGC,Finkelman2011Hoag,MutluPakdil2017}. Published component magnitudes and structural parameters are not always directly comparable because the studies use different imaging depths, passbands, spatial resolutions, component models, and integration apertures. Some earlier studies measured ring fluxes with the central region excluded, whereas \citet{MutluPakdil2017}, like the present work, used core-subtracted residual images. Core subtraction can reduce contamination of the ring flux by underlying core light, but differences in the reported parameters cannot be attributed to this methodological choice alone. For example, we obtain higher integrated core luminosities but smaller $R_{\rm e}$ for Hoag's Object and UGC 4599 than reported by \citet{Finkelman2011Hoag} and \citet{Finkelman2011UGC}. Different photometric definitions may contribute to these cross-study offsets: those studies measured core fluxes within the observed core--ring boundaries and obtained effective radii from fitted de Vaucouleurs or S\'ersic profiles, whereas we consistently integrate the reconstructed core profiles within $2R_{\rm P}$ and determine $R_{\rm e}$ from half of that enclosed light for our sample. A finite-aperture half-light radius need not equal the effective radius of a model integrated to infinity. The relative contributions of these methodological differences to the luminosity and radius offsets remain undetermined. 

The analyses also differ in scope: our broad-band data do not test previous constraints on stellar ages or gas kinematics, including evidence for old cores, younger star-forming rings, and extended H\,{\sc i} structures around Hoag's Object and UGC 4599 \citep{Schweizer1987,Finkelman2011UGC,Finkelman2011Hoag,Brosch2013}. For PGC 1000714, we measure the main detached outer ring without separately isolating its additional red inner ring \citep{MutluPakdil2017}. For UGC 4599, the deeper imaging of \citet{Mosenkov2023} reveals a giant low-surface-brightness spiral system rather than a genuinely detached ring; we discuss this classification issue in Section~\ref{subsec:sample_purity}. 

The main advantage of the present work is that a substantially larger sample is analysed with uniform imaging, component definitions, and photometric procedures. This removes much of the methodological heterogeneity among individual studies and allows both the diversity and the common statistical properties of Hoag-type systems to be quantified. In particular, the rings are not uniformly blue but extend to the red sequence, while the cores show slightly different properties compared with mass-matched early-type galaxies. The high central fraction and the correlations between ring structure and the core-plus-inner-halo to ring mass ratio likewise emerge only at the population level. These results place the few well-studied objects within a broader distribution and show that no single system provides a complete template for the class.

\subsection{Implications for formation and evolution}
\label{subsec:formation_implications}

The slightly bluer core colours, lower concentrations, and lower S\'ersic indices suggest that secular evolution may have contributed to some Hoag-type cores. Pseudobulges are bluer on average than classical bulges \citep{Gadotti2009}; \citet{FisherDrory2008} found that they typically have $n<2$ and linked their structure to secular processes, while \citet{KormendyKennicutt2004} reviewed how bars and other non-axisymmetric structures redistribute disc material. However, as discussed in Section~\ref{subsec:core_scaling}, the higher stellar-mass surface densities and higher mean effective surface brightnesses of the cores distinguish them from the typical diffuse pseudobulge population. Their structural assembly histories may therefore be more complex than a simple secular origin for the entire core.

Their predominance among central galaxies may be more directly related to the persistence of the Hoag-type morphology than to a unique core formation mechanism. Satellite galaxies are preferentially affected by environmental quenching after infall into a larger halo \citep{vanDenBosch2008,Peng2012,Wetzel2013}. This predominance of centrals may help these systems retain the gas reservoirs that sustain star formation in the outer rings and reduce environmental damage to these low-density structures. The high central fraction could therefore partly reflect a greater probability of ring survival, allowing systems with different formation histories to remain recognizable as Hoag-type galaxies for longer.

Studies of individual Hoag-type galaxies further illustrate how different formation pathways may lead to similar present-day morphologies. More importantly, detailed observations identify the core of Hoag's Object as an old, rapidly rotating but genuinely elliptical and mildly triaxial system without a stellar disc or bar, arguing against the dissolved-bar pathway for the prototype \citep{Finkelman2011Hoag}. Gas accretion onto a pre-existing spheroid remains compatible with a red core, a younger ring, isolation, and an extended H\,{\sc i} reservoir, while older mergers or accretion events cannot be excluded when tidal signatures have faded \citep{Schweizer1987,Finkelman2011UGC,Brosch2013}; a recent accretion event has also been proposed for PGC 1000714 \citep{MutluPakdil2017}. The wide distributions of ring colour and $\eta$ are consistent with multiple formation pathways, or with variations in initial conditions and evolutionary stage within a common pathway.

\subsection{Sample purity}
\label{subsec:sample_purity}

The purity of a visually selected Hoag-type sample depends on the depth at which the separation between the core and ring can be assessed. UGC 4599 provides a concrete example. Although its bright central body and surrounding ring-like structure resemble a Hoag-type galaxy in the DESI images, the deeper observations of \citet{Mosenkov2023} reveal spiral structure extending from the central region through the apparent ring, embedded in a giant low-surface-brightness (gLSB) disc. Its central component is also distinct in the scaling relations: it departs markedly towards lower surface brightness from the elliptical/classical-bulge Kormendy relation, whereas the core of Hoag's Object lies within the elliptical-galaxy locus \citep{Mosenkov2023}. Our measurements likewise place UGC 4599 on the low-surface-brightness side of the early-type comparison population in Figure~\ref{fig:core_sb_radius}. Its core is also bluer and less massive than most of the other cores in our sample (Figures~\ref{fig:mag_colour} and \ref{fig:core_mass_params}; Table~\ref{tab:core_properties}). Both its spiral structure and its position in the scaling relations therefore distinguish UGC 4599 from Hoag's Object and suggest a different structural assembly history. 

The DESI images (Figure~\ref{fig:sample_examples}) show an uneven elliptical ring around LEDA 4547778. The outer structure of SDSS J125253.45+531427.1 is more diffuse and asymmetric, with uncertain closure. We do not identify a reliable continuous spiral connection between the core and ring in either object, so these images alone do not establish them as spiral contaminants.

Repeating the comparisons in Figure~\ref{fig:core_params_1} after excluding UGC 4599, or the three core outliers identified in Section~\ref{subsec:photometric_distributions}, preserves the tendencies towards bluer colours, higher stellar-mass surface densities, and lower concentrations and S\'ersic indices. The colour, surface-density, concentration, and S\'ersic-index differences retain $p\leq7.9\times10^{-3}$ in both tests, whereas the size difference remains statistically insignificant ($p=0.21$ and $0.18$, respectively). Thus, the core-property differences shown in Figure~\ref{fig:core_params_1} are not driven solely by these objects.

The ring correlations in Figure~\ref{fig:mass_ratio_ring_params} also persist after excluding UGC 4599 or the three outliers. For these two exclusions, respectively, the Spearman coefficients are $\rho=-0.82\pm0.07$ and $-0.81\pm0.08$ for ring stellar-mass surface density, and $\rho=-0.77\pm0.10$ and $-0.78\pm0.10$ for normalized ring width. All four tests give $p\leq1.6\times10^{-6}$. These two anticorrelations are therefore not driven solely by the excluded objects.

The apparent ring of UGC 4599 is part of a spiral pattern rather than a detached annulus separated from the core by a genuine gap, making it a contaminant under our adopted morphological definition. Its inclusion reflects the insufficient depth of our DESI images to reveal these faint connections. Its reclassification indicates that other galaxies in the sample may also contain faint connecting spiral arms or discs that remain undetected in the available imaging. The current sample should therefore be regarded as selected by apparent Hoag-type morphology at the DESI imaging depth, rather than as a fully confirmed set of intrinsically detached core--ring systems. Deeper imaging is needed to establish whether the gaps remain distinct at lower surface brightness, identify similar contaminants, and assess the sample purity.

\subsection{Limitations and prospects}
\label{subsec:limitations}

The sample comprises only thirty visually selected systems, with SDSS-based group information available for sixteen. Its completeness and purity are therefore not fully established, and the inferred central fraction remains subject to limited statistics and group-catalogue coverage. As discussed in Section~\ref{subsec:sample_purity}, deeper imaging is needed to distinguish detached rings from faint spiral or disc connections. A larger sample selected with consistent criteria and comparable imaging depth, together with more complete environmental information, would test whether the core-property offsets and high central fraction are representative of the wider population.

The broad-band photometry constrains component colours and structures but cannot uniquely separate stellar age, metallicity, and dust effects, establish a pseudobulge classification, or identify the origin of the ring gas. Present-day central--satellite classifications likewise do not recover past interactions or demonstrate that central status causes longer ring survival. Spatially resolved spectroscopy would constrain the stellar populations and relative rotation of the cores and rings, while H\,{\sc i} and CO observations would measure the gas reservoirs and their kinematics. Together with deep imaging for embedded structures and tidal debris, these data would help distinguish secular evolution, external accretion, and subsequent environmental processing.

\section{Conclusions}
\label{sec:summary}

We identified thirty galaxies with apparent Hoag-type morphology in DESI imaging by visually inspecting two machine-learning ring-galaxy catalogues \citep{Abraham2025,Zhang2025}, extending previous studies of individual systems to a statistical sample. Using DESI Legacy Imaging Surveys $g$, $r$, and $z$ images, we performed homogeneous photometry of the cores and rings. We fitted the cores with the rings masked, subtracted the resulting two-dimensional core models, and measured the rings from the residual images to derive the colours, stellar masses, and structural properties of both components. We compared the cores with SDSS early-type galaxies matched in stellar mass and, for environmental comparisons, central--satellite status. We also examined the relation between ring structure and the core-plus-inner-halo to ring mass ratio and compared the component radii with a simplified orbital-stability model.

Our main conclusions are as follows:
\begin{enumerate}
    \item The cores lie predominantly on the red sequence, whereas the rings span a broader colour range extending toward the blue cloud.
    \item The cores broadly follow early-type mass scaling relations, but compared with the mass-matched early-type sample, they tend to be slightly bluer and smaller, with higher stellar-mass surface densities and lower concentrations and single-component S\'ersic indices.
    \item Fourteen of the 16 systems with SDSS group information are centrals (87.5 per cent), compared with 63.75 per cent in SampleC. The halo-mass and overdensity distributions become much more similar after central--satellite status is also matched.
    \item Ring stellar-mass surface density and normalized width decrease with increasing $\eta=(M_{\rm core}+M_{\rm DM})/M_{\rm ring}$.
    \item Most core $R_{90}$ values lie inside the OSCO of the simplified stability model, while the observed rings frequently extend across the OSCO--LC interval.
\end{enumerate}

\section*{Acknowledgements}
J.Y. thanks Yanhan Guo and Ce Gao for helpful discussions. This work is supported by the National Natural Science Foundation of China (grant No. 12433003), and the National Key R\&D Program of China (grant No. 2018YFA0404502).

Funding for the SDSS and SDSS-II has been provided by the Alfred P. Sloan Foundation, the Participating Institutions, the National Science Foundation, the U.S. Department of Energy, the National Aeronautics and Space Administration, the Japanese Monbukagakusho, the Max Planck Society, and the Higher Education Funding Council for England. The SDSS Web Site is \url{www.sdss.org}.

The SDSS is managed by the Astrophysical Research Consortium for the Participating Institutions. The Participating Institutions are the American Museum of Natural History, Astrophysical Institute Potsdam, University of Basel, University of Cambridge, Case Western Reserve University, University of Chicago, Drexel University, Fermilab, the Institute for Advanced Study, the Japan Participation Group, Johns Hopkins University, the Joint Institute for Nuclear Astrophysics, the Kavli Institute for Particle Astrophysics and Cosmology, the Korean Scientist Group, the Chinese Academy of Sciences (LAMOST), Los Alamos National Laboratory, the Max-Planck-Institute for Astronomy (MPIA), the Max-Planck-Institute for Astrophysics (MPA), New Mexico State University, Ohio State University, University of Pittsburgh, University of Portsmouth, Princeton University, the United States Naval Observatory, and the University of Washington.

The Legacy Surveys consist of three individual and complementary projects: the Dark Energy Camera Legacy Survey (DECaLS; Proposal ID \#2014B-0404; PIs: David Schlegel and Arjun Dey), the Beijing-Arizona Sky Survey (BASS; NOAO Prop. ID \#2015A-0801; PIs: Zhou Xu and Xiaohui Fan), and the Mayall z-band Legacy Survey (MzLS; Prop. ID \#2016A-0453; PI: Arjun Dey). DECaLS, BASS and MzLS together include data obtained, respectively, at the Blanco telescope, Cerro Tololo Inter-American Observatory, NSF’s NOIRLab; the Bok telescope, Steward Observatory, University of Arizona; and the Mayall telescope, Kitt Peak National Observatory, NOIRLab. Pipeline processing and analyses of the data were supported by NOIRLab and the Lawrence Berkeley National Laboratory (LBNL). The Legacy Surveys project is honored to be permitted to conduct astronomical research on I’oligam Du’ag (Kitt Peak), a mountain with particular significance to the \href{https://www.tonation-nsn.gov/}{Tohono O’odham Nation}.

NOIRLab is operated by the Association of Universities for Research in Astronomy (AURA) under a cooperative agreement with the National Science Foundation. LBNL is managed by the Regents of the University of California under contract to the U.S. Department of Energy.

This project used data obtained with the Dark Energy Camera (DECam), which was constructed by the Dark Energy Survey (DES) collaboration. Funding for the DES Projects has been provided by the U.S. Department of Energy, the U.S. National Science Foundation, the Ministry of Science and Education of Spain, the Science and Technology Facilities Council of the United Kingdom, the Higher Education Funding Council for England, the National Center for Supercomputing Applications at the University of Illinois at Urbana-Champaign, the Kavli Institute of Cosmological Physics at the University of Chicago, Center for Cosmology and Astro-Particle Physics at the Ohio State University, the Mitchell Institute for Fundamental Physics and Astronomy at Texas A\&M University, Financiadora de Estudos e Projetos, Fundacao Carlos Chagas Filho de Amparo a Pesquisa do Estado do Rio de Janeiro, Conselho Nacional de Desenvolvimento Cientifico e Tecnologico and the Ministerio da Ciencia, Tecnologia e Inovacao, the Deutsche Forschungsgemeinschaft and the Collaborating Institutions in the Dark Energy Survey. The Collaborating Institutions are Argonne National Laboratory, the University of California at Santa Cruz, the University of Cambridge, Centro de Investigaciones Energeticas, Medioambientales y Tecnologicas-Madrid, the University of Chicago, University College London, the DES-Brazil Consortium, the University of Edinburgh, the Eidgenossische Technische Hochschule (ETH) Zurich, Fermi National Accelerator Laboratory, the University of Illinois at Urbana-Champaign, the Institut de Ciencies de l’Espai (IEEC/CSIC), the Institut de Fisica d’Altes Energies, Lawrence Berkeley National Laboratory, the Ludwig Maximilians Universitat Munchen and the associated Excellence Cluster Universe, the University of Michigan, NSF’s NOIRLab, the University of Nottingham, the Ohio State University, the University of Pennsylvania, the University of Portsmouth, SLAC National Accelerator Laboratory, Stanford University, the University of Sussex, and Texas A\&M University.

BASS is a key project of the Telescope Access Program (TAP), which has been funded by the National Astronomical Observatories of China, the Chinese Academy of Sciences (the Strategic Priority Research Program “The Emergence of Cosmological Structures” Grant \# XDB09000000), and the Special Fund for Astronomy from the Ministry of Finance. The BASS is also supported by the External Cooperation Program of Chinese Academy of Sciences (Grant \# 114A11KYSB20160057), and Chinese National Natural Science Foundation (Grant \# 12120101003, \# 11433005).

The Legacy Survey team makes use of data products from the Near-Earth Object Wide-field Infrared Survey Explorer (NEOWISE), which is a project of the Jet Propulsion Laboratory/California Institute of Technology. NEOWISE is funded by the National Aeronautics and Space Administration.

The Legacy Surveys imaging of the DESI footprint is supported by the Director, Office of Science, Office of High Energy Physics of the U.S. Department of Energy under Contract No. DE-AC02-05CH11231, by the National Energy Research Scientific Computing Center, a DOE Office of Science User Facility under the same contract; and by the U.S. National Science Foundation, Division of Astronomical Sciences under Contract No. AST-0950945 to NOAO.

\section*{Data Availability}

All the data used in this work are publicly available, including the following: (1) ring galaxy catalogues from \citet{Abraham2025} and \citet{Zhang2025}; (2) imaging data from the DESI Legacy Imaging Surveys DR9 \citep{Dey2019}; (3) the New York University Value-Added Galaxy Catalog \citep[NYU-VAGC;][]{Blanton2005}; (4) the NASA-Sloan Atlas \citep[NSA;][]{Blanton2011}; (5) the SDSS DR7 group catalogue from \citet{Yang2007}; (6) the ELUCID environmental data \citep{Wang2016ELUCID}; (7) the DESI group catalogues from \citet{Yang2021} and \citet{Williams2026}.

The derived core and ring photometric properties are provided in Tables~\ref{tab:core_properties} and~\ref{tab:ring_properties}. Additional data products are available from the corresponding authors upon reasonable request.

\bibliographystyle{mnras}
\bibliography{references}

\appendix
\section{Individual galaxy atlas}
\label{app:galaxy_atlas}

Figures~\ref{fig:appendix_galaxy_atlas_1}--\ref{fig:appendix_galaxy_atlas_6} show the individual decomposition figures for the thirty Hoag-type galaxies. The labels above each panel are the catalogue names listed in Table~\ref{tab:core_properties}; the panel format follows Figure~\ref{fig:decomposition_example}.

\newcommand{\atlaspanel}[2]{%
    {\small\bfseries #1}\par\nobreak
    \includegraphics[width=0.92\textwidth]{figures/figure_of_each_galaxy/#2}\par\vspace{0.25em}%
}

\begin{figure*}
    \centering
    \atlaspanel{SDSS J004228.79+161728.5}{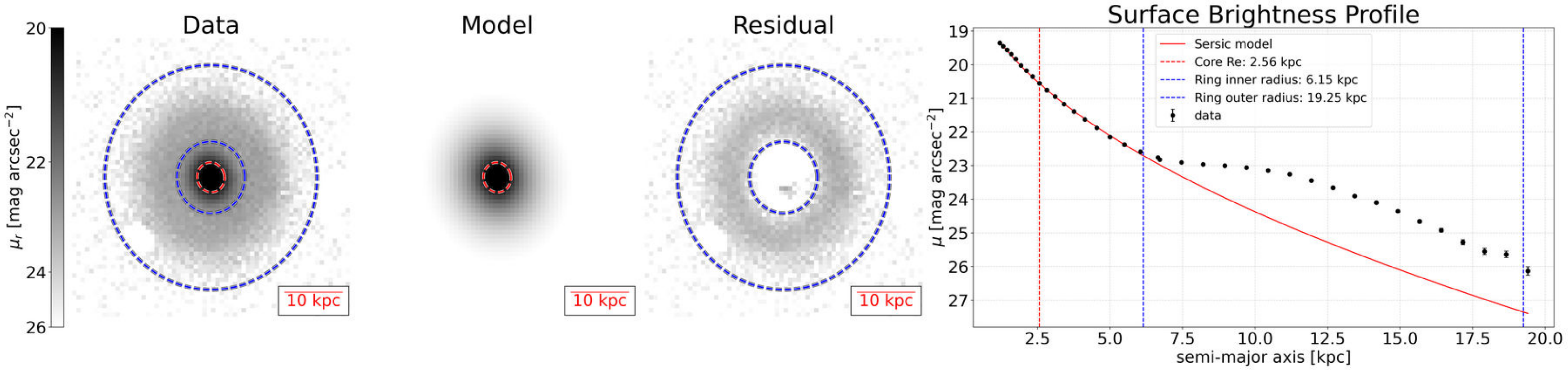}
    \atlaspanel{LEDA 1818512}{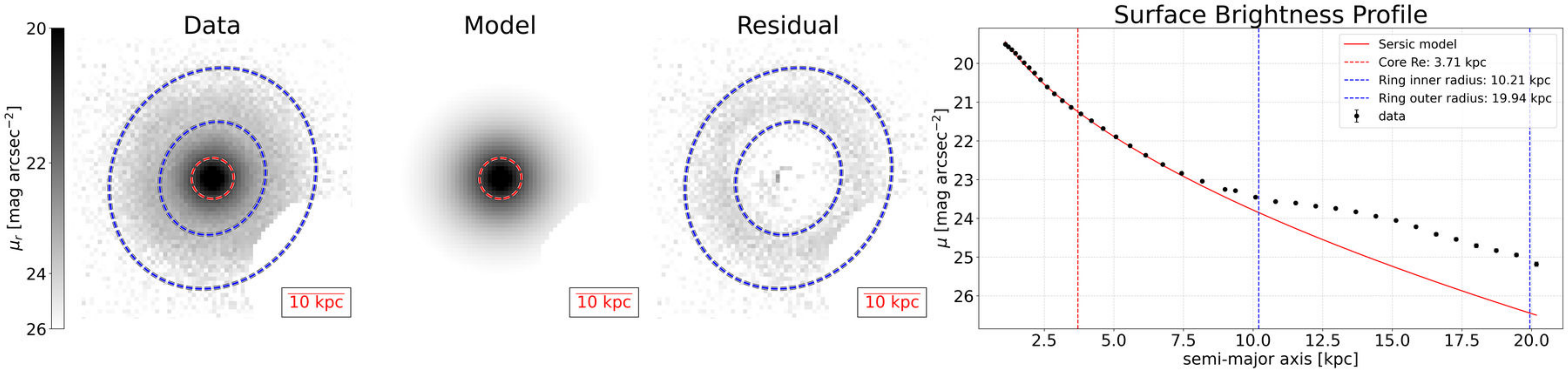}
    \atlaspanel{SDSS J082823.59+015614.6}{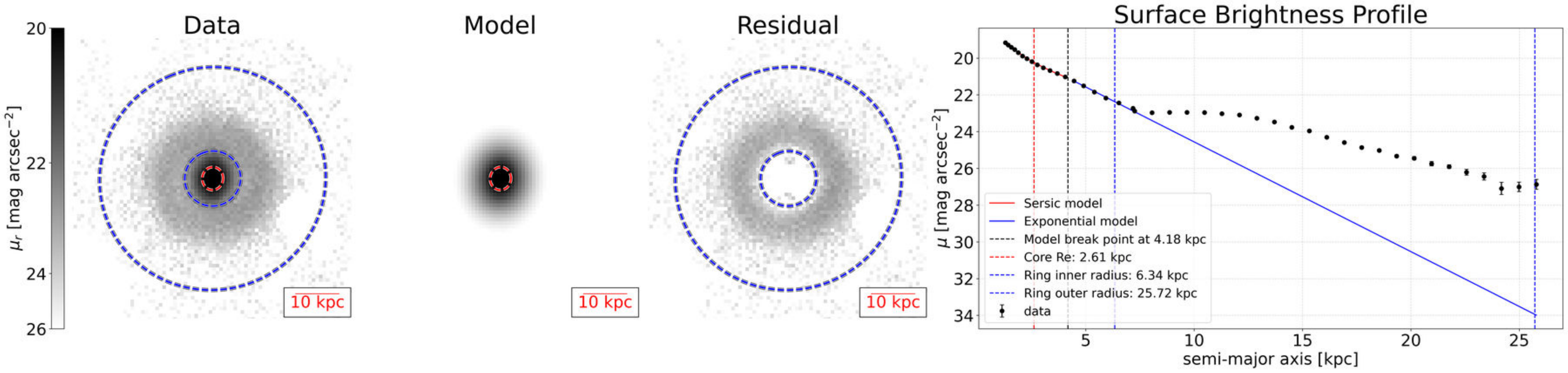}
    \atlaspanel{UGC 4599}{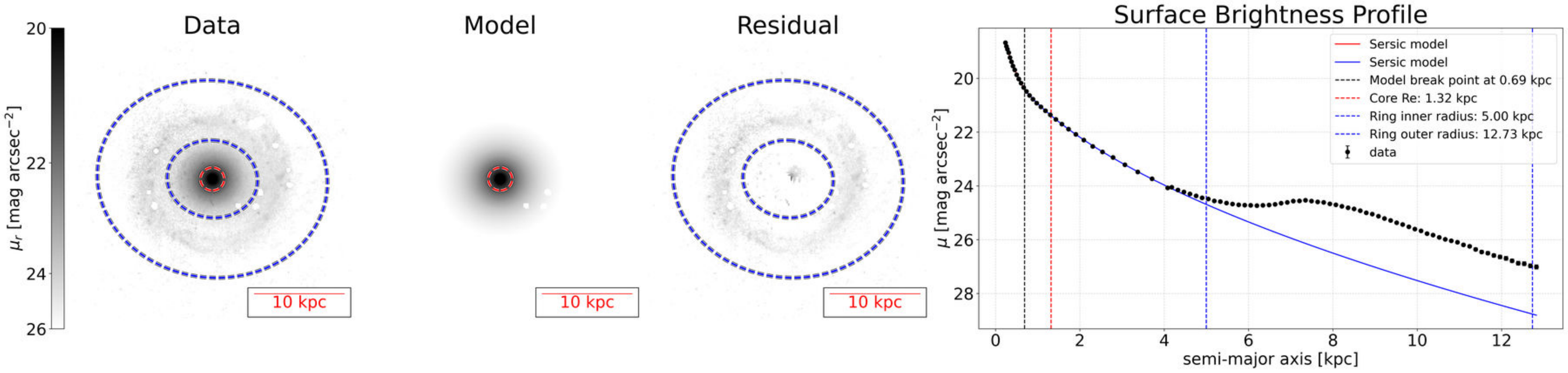}
    \atlaspanel{LEDA 1668413}{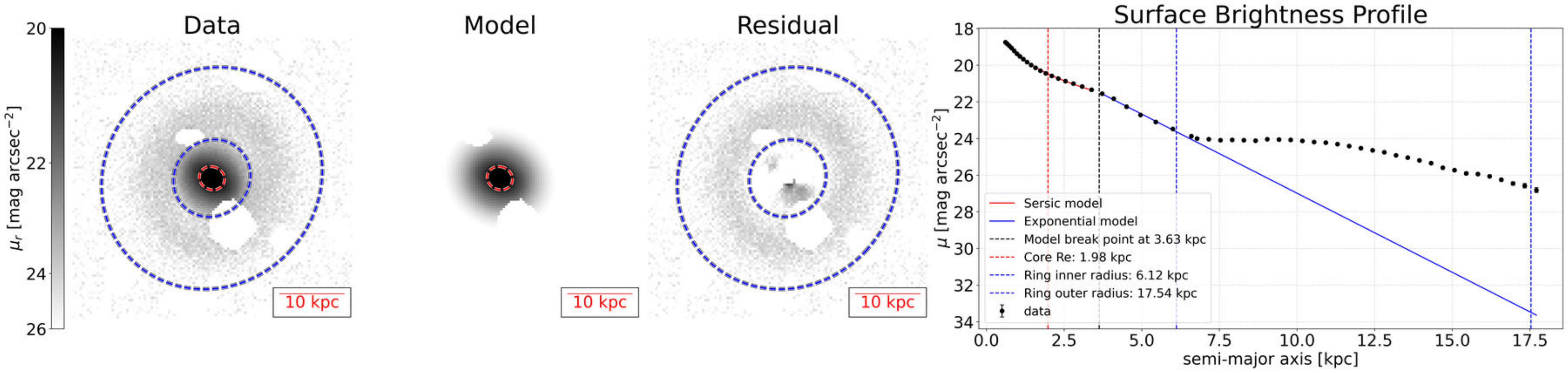}
    \caption{Individual galaxy decomposition figures.}
    \label{fig:appendix_galaxy_atlas_1}
\end{figure*}

\begin{figure*}
    \centering
    \atlaspanel{LEDA 3097463}{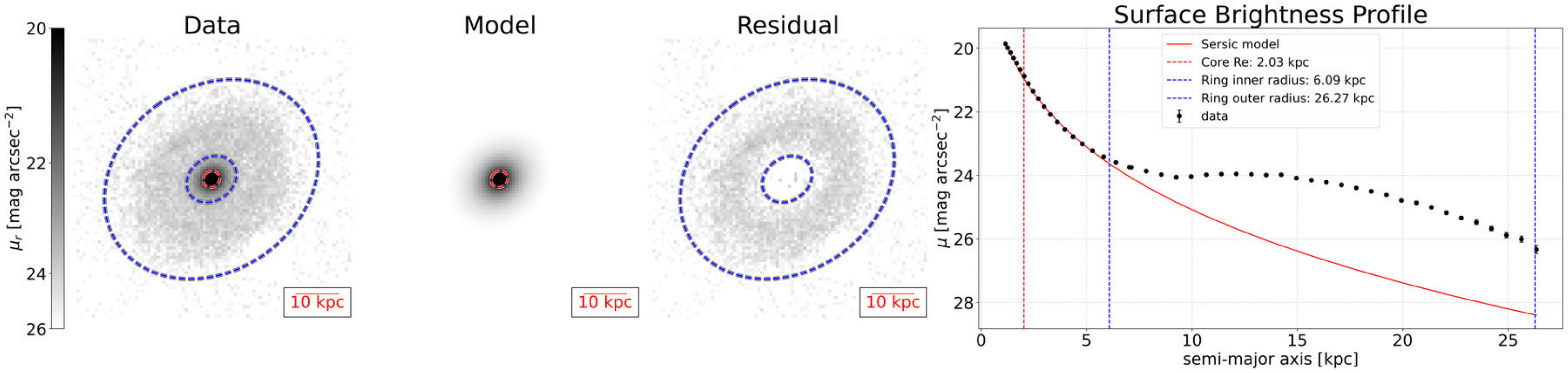}
    \atlaspanel{LEDA 1829531}{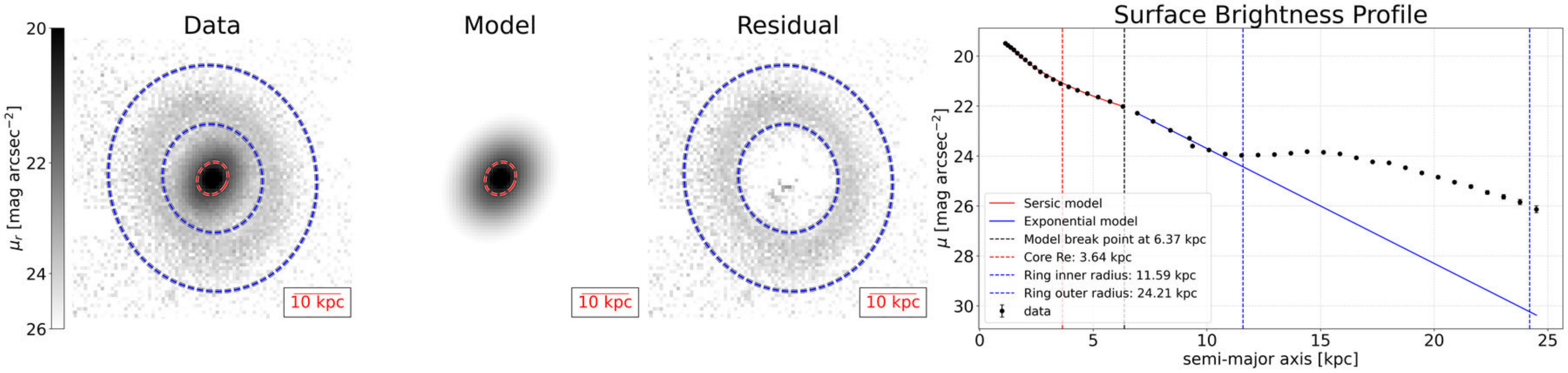}
    \atlaspanel{LEDA 1564718}{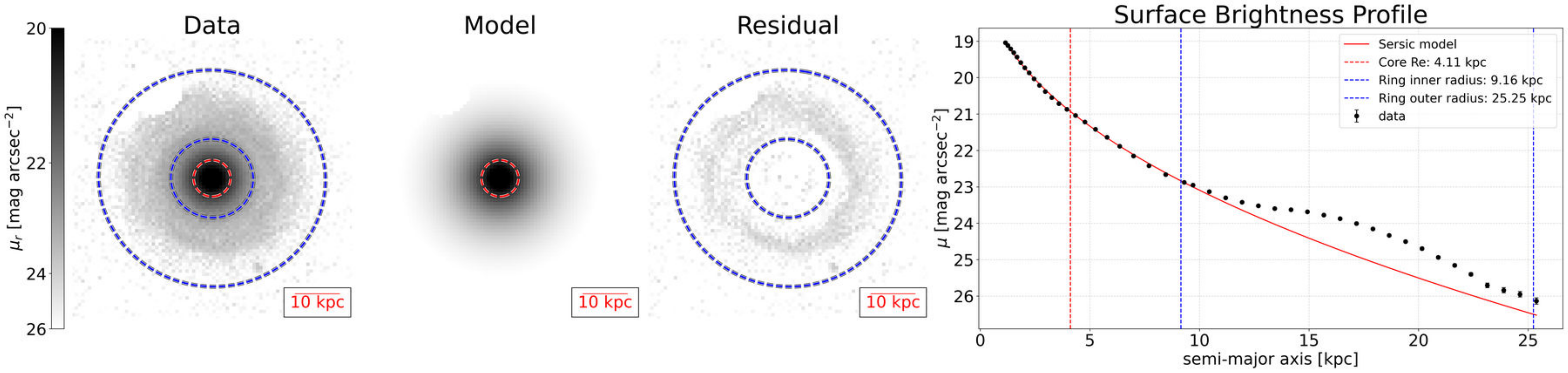}
    \atlaspanel{SDSS J010054.43+253637.1}{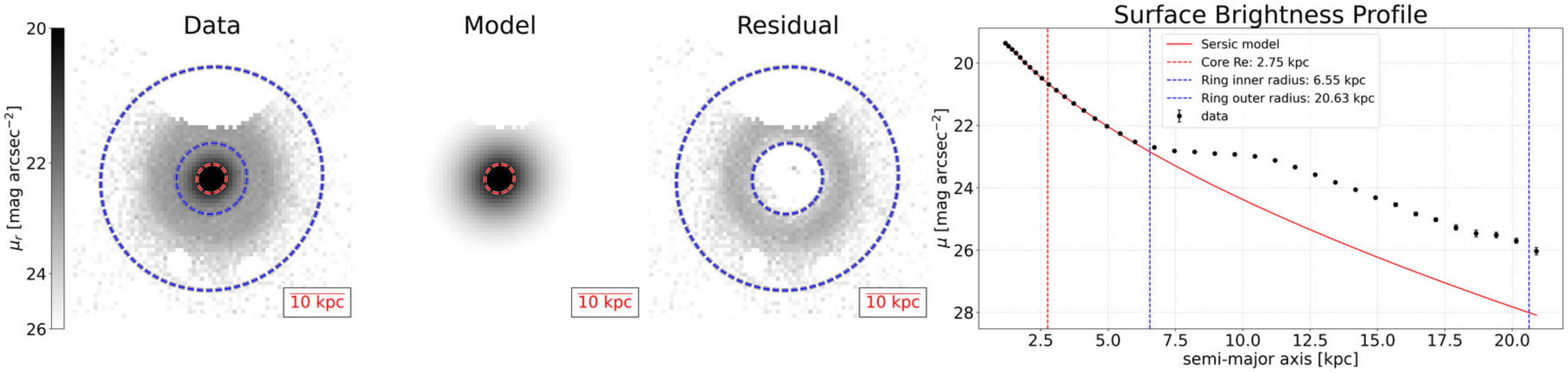}
    \atlaspanel{LEDA 1840343}{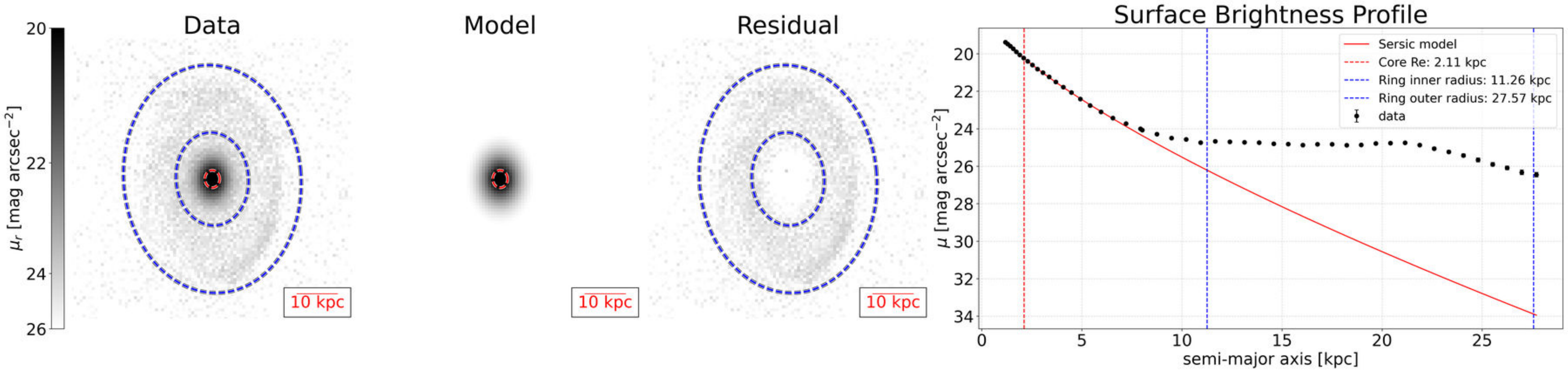}
    \caption{Individual galaxy decomposition figures (continued).}
    \label{fig:appendix_galaxy_atlas_2}
\end{figure*}

\begin{figure*}
    \centering
    \atlaspanel{LEDA 986443}{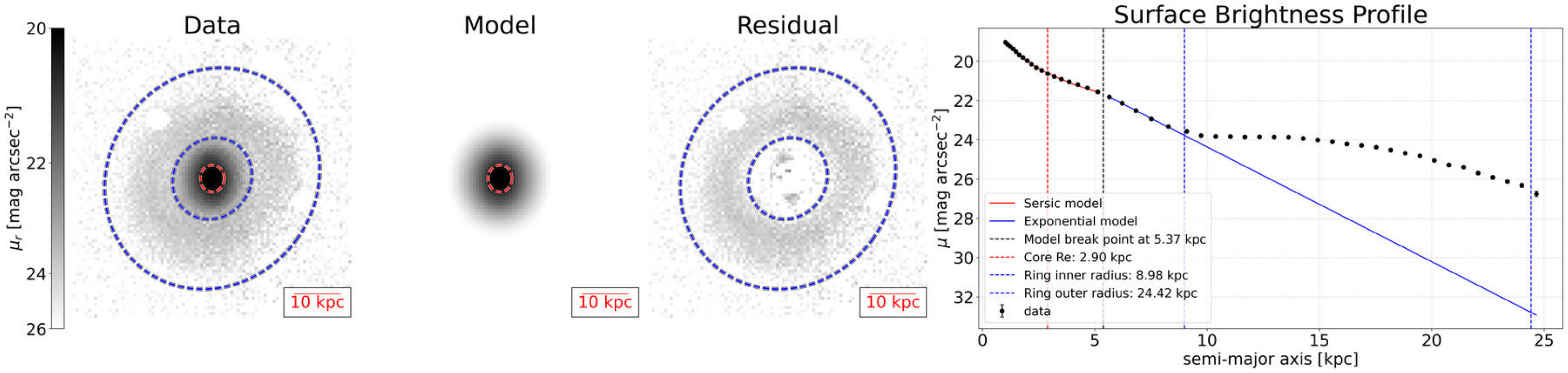}
    \atlaspanel{PGC 1000714}{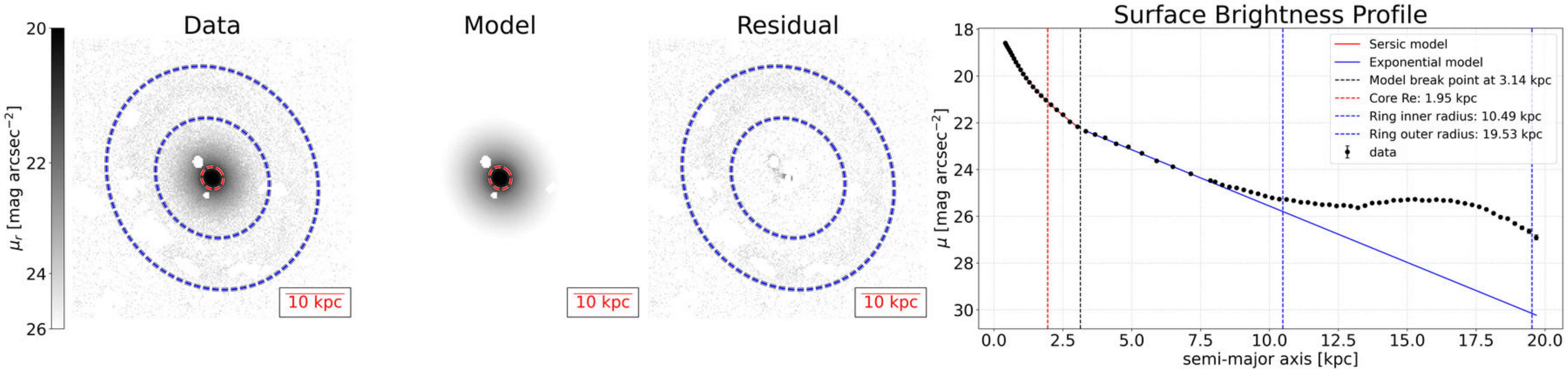}
    \atlaspanel{Z 39-164}{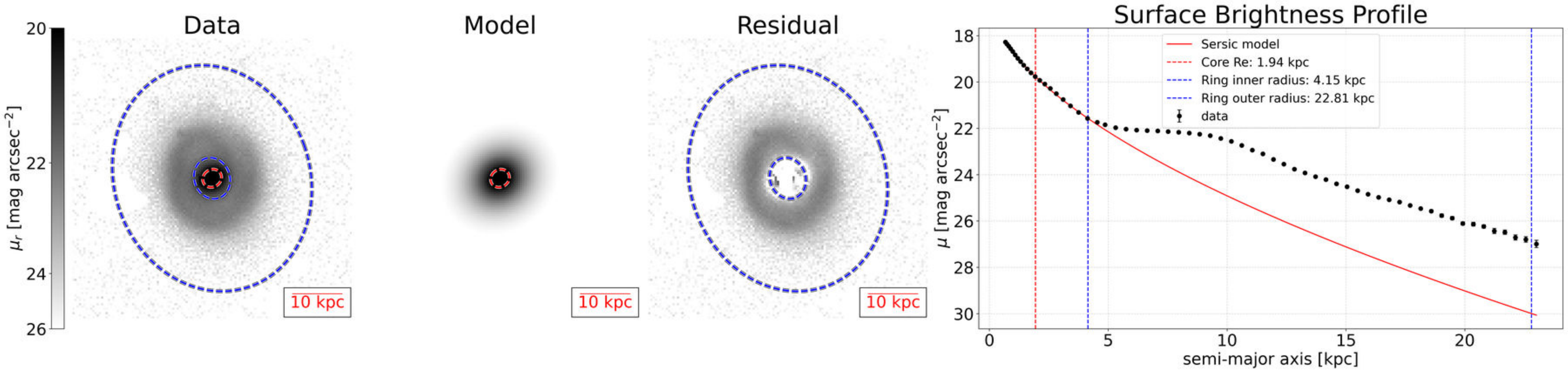}
    \atlaspanel{LEDA 4547778}{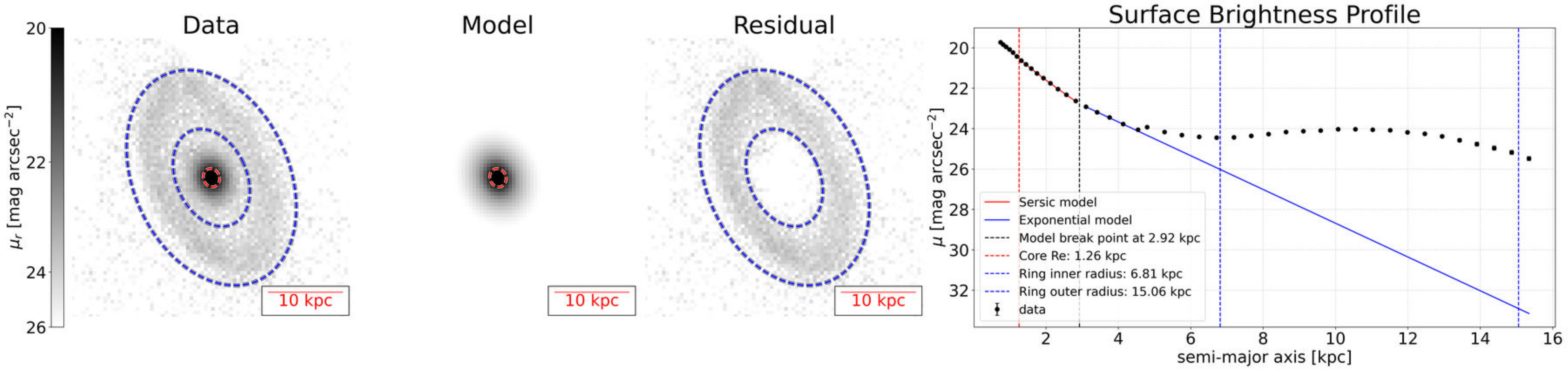}
    \atlaspanel{2MASS J12014342+0010593}{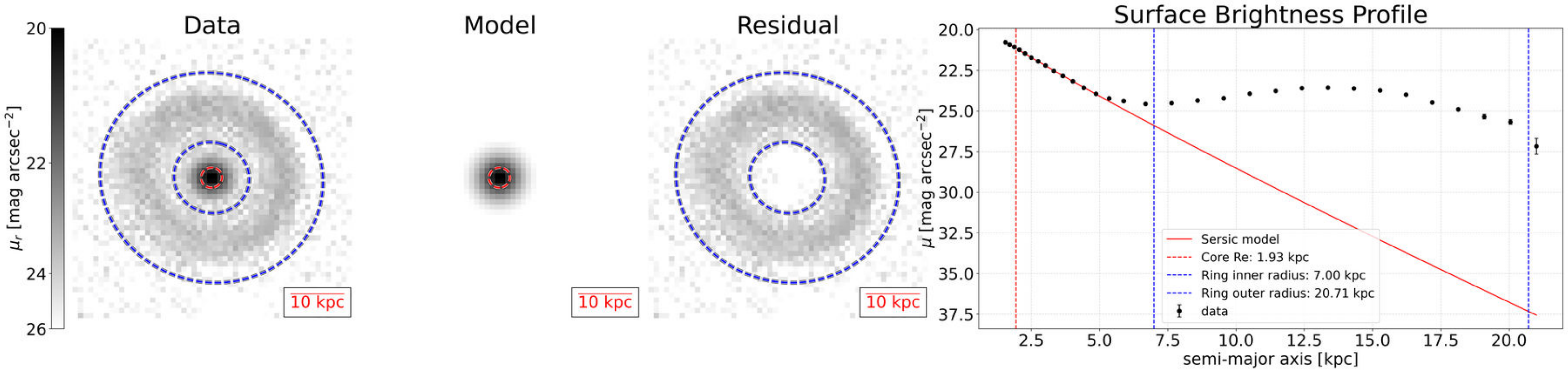}
    \caption{Individual galaxy decomposition figures (continued).}
    \label{fig:appendix_galaxy_atlas_3}
\end{figure*}

\begin{figure*}
    \centering
    \atlaspanel{2MASX J12363134+2325332}{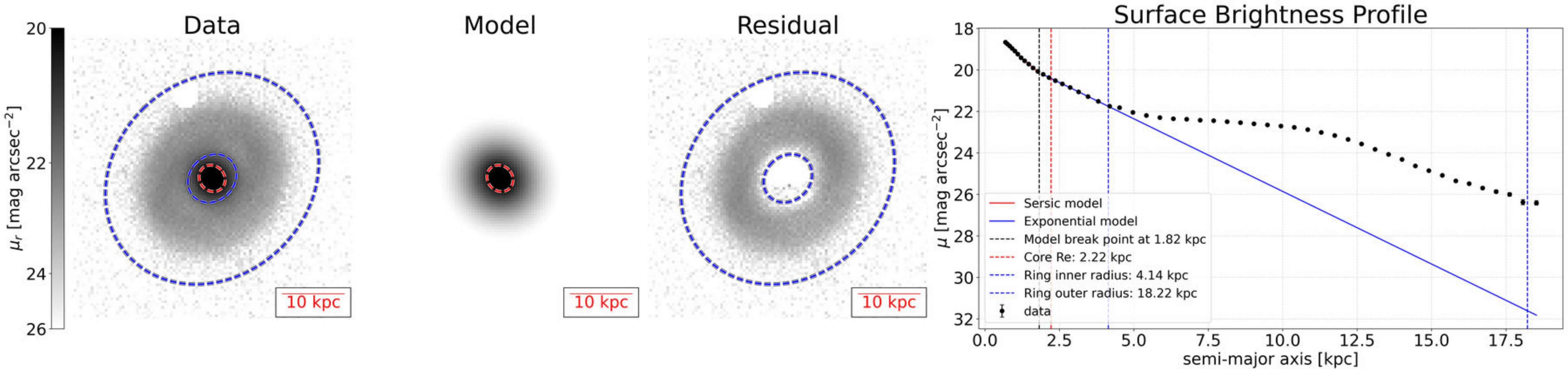}
    \atlaspanel{Z 70-200}{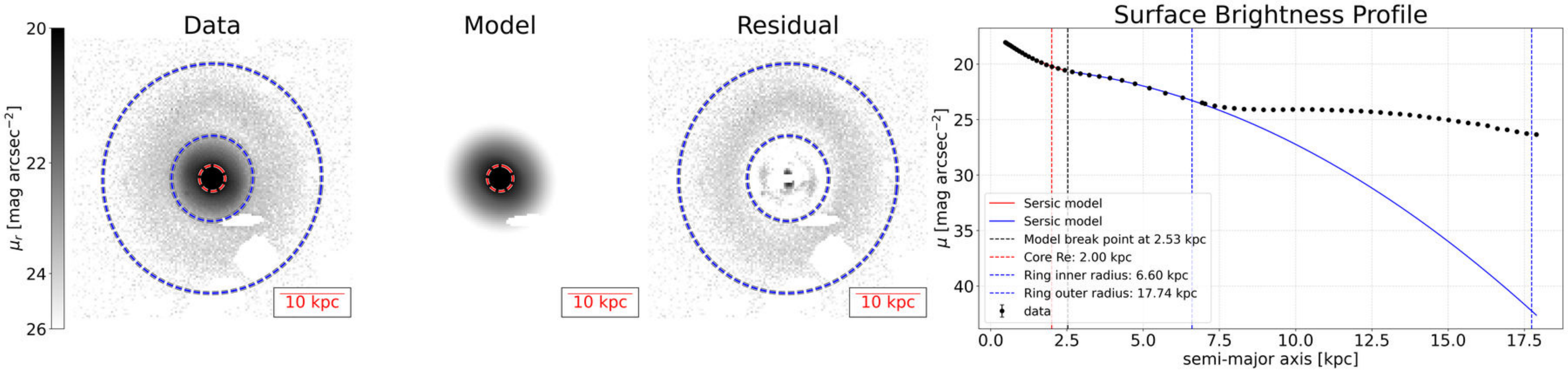}
    \atlaspanel{SDSS J125253.45+531427.1}{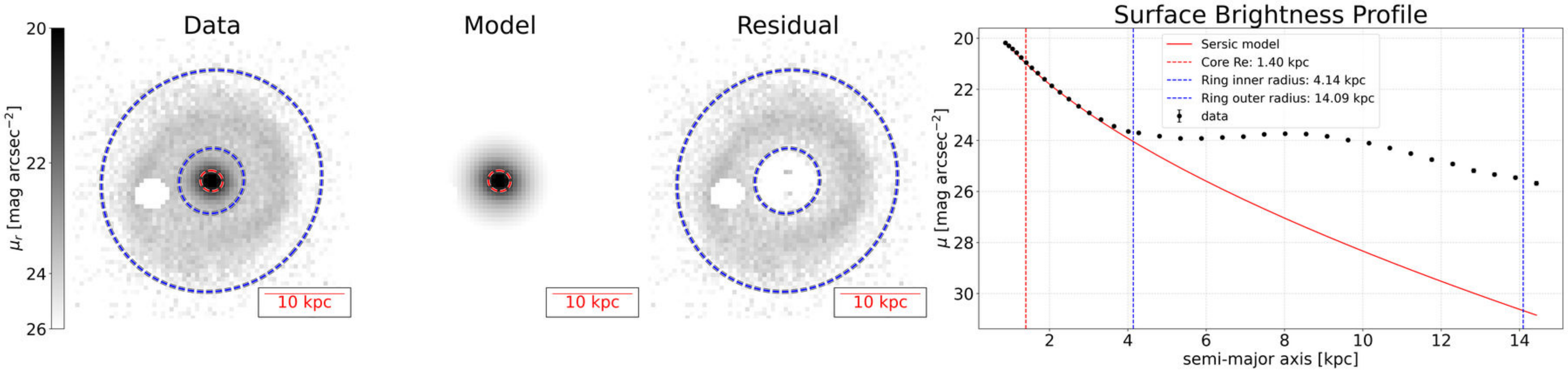}
    \atlaspanel{2MASX J13130376+1050115}{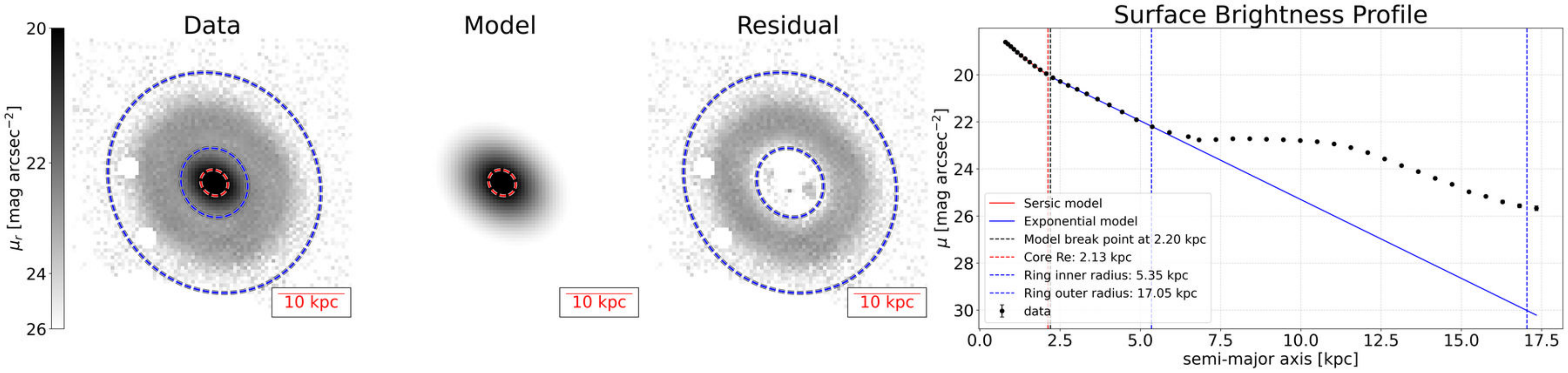}
    \atlaspanel{SDSS J011843.87+250129.9}{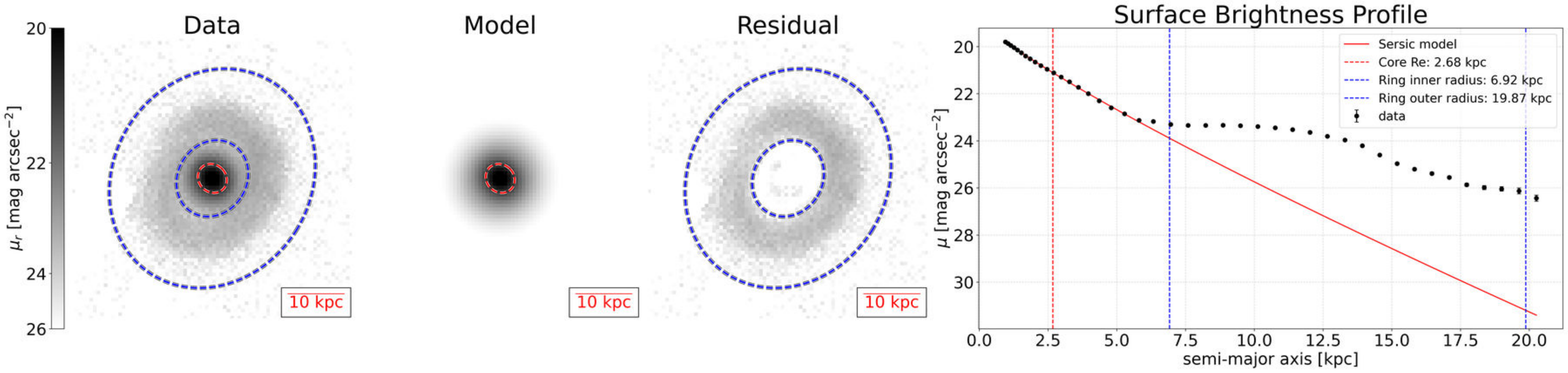}
    \caption{Individual galaxy decomposition figures (continued).}
    \label{fig:appendix_galaxy_atlas_4}
\end{figure*}

\begin{figure*}
    \centering
    \atlaspanel{SDSSCGB 8233.1}{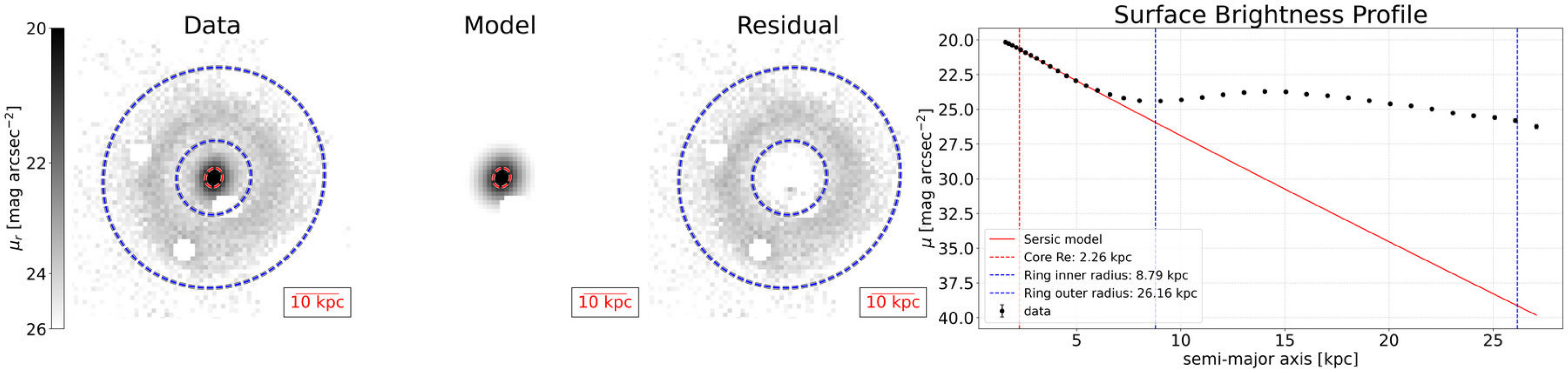}
    \atlaspanel{SDSS J132549.22-015957.5}{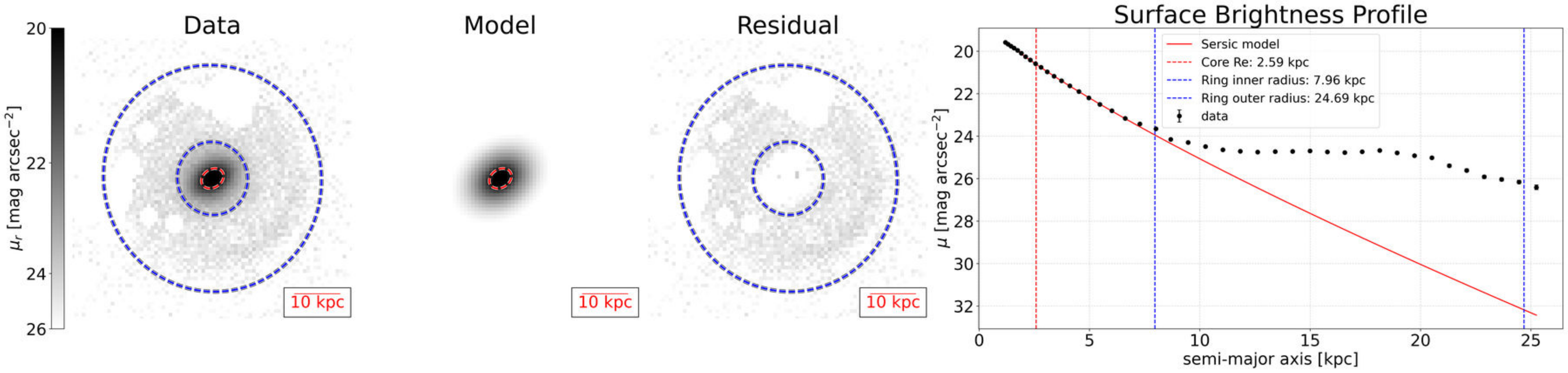}
    \atlaspanel{ZW VIII 437}{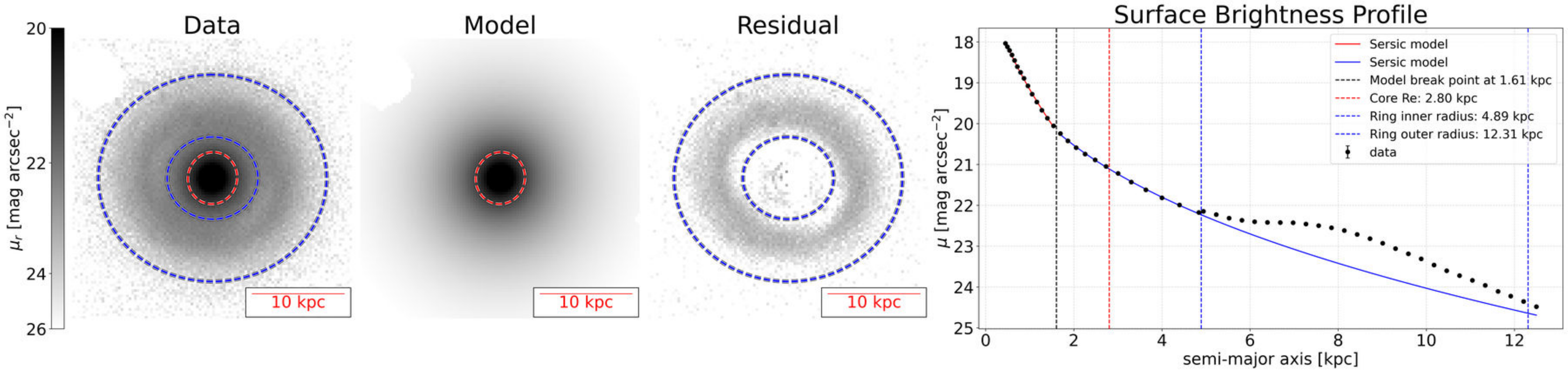}
    \atlaspanel{WISEA J151059.53-043028.0}{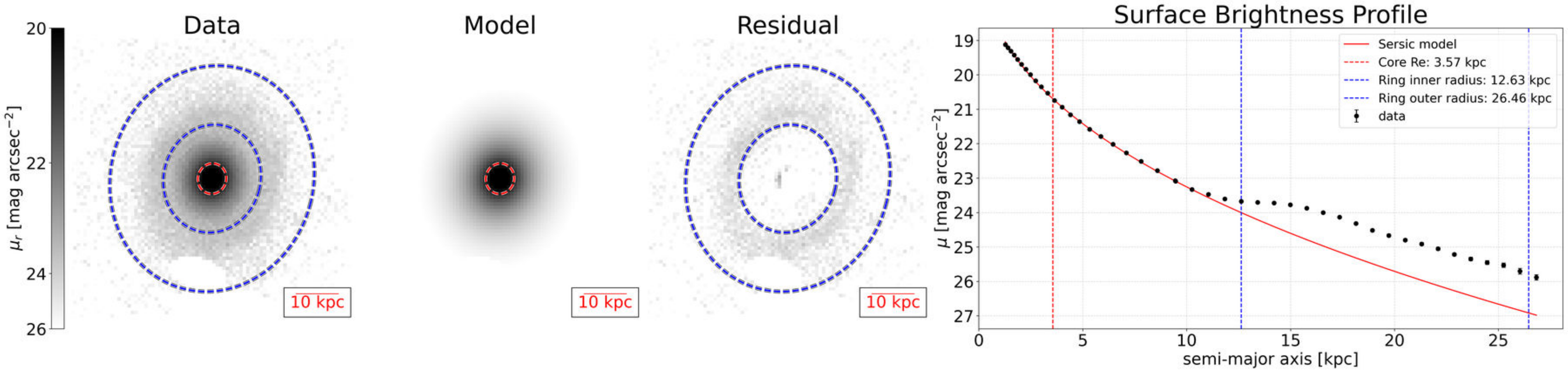}
    \atlaspanel{Hoag's Object}{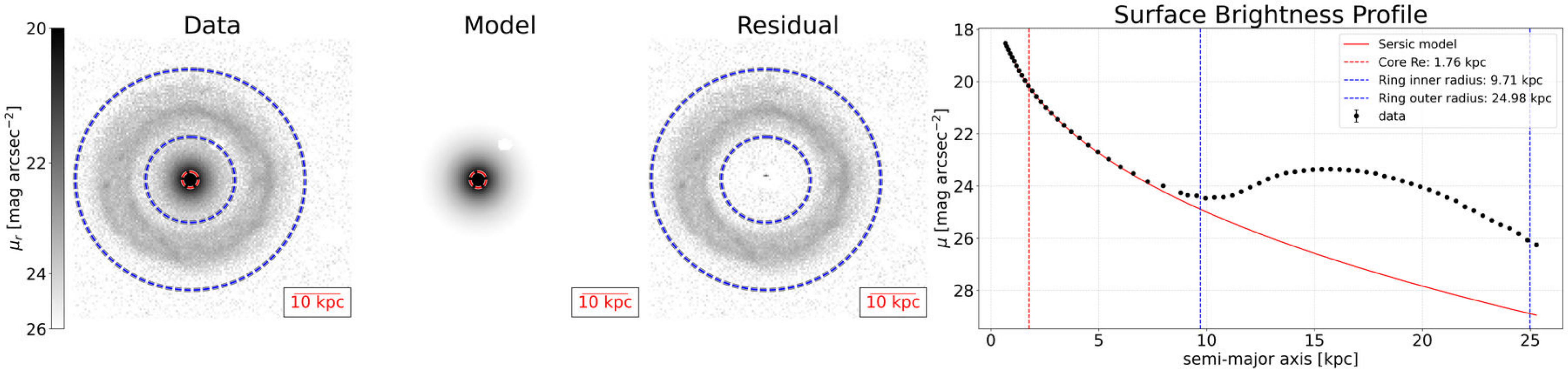}
    \caption{Individual galaxy decomposition figures (continued).}
    \label{fig:appendix_galaxy_atlas_5}
\end{figure*}

\begin{figure*}
    \centering
    \atlaspanel{LEDA 1543950}{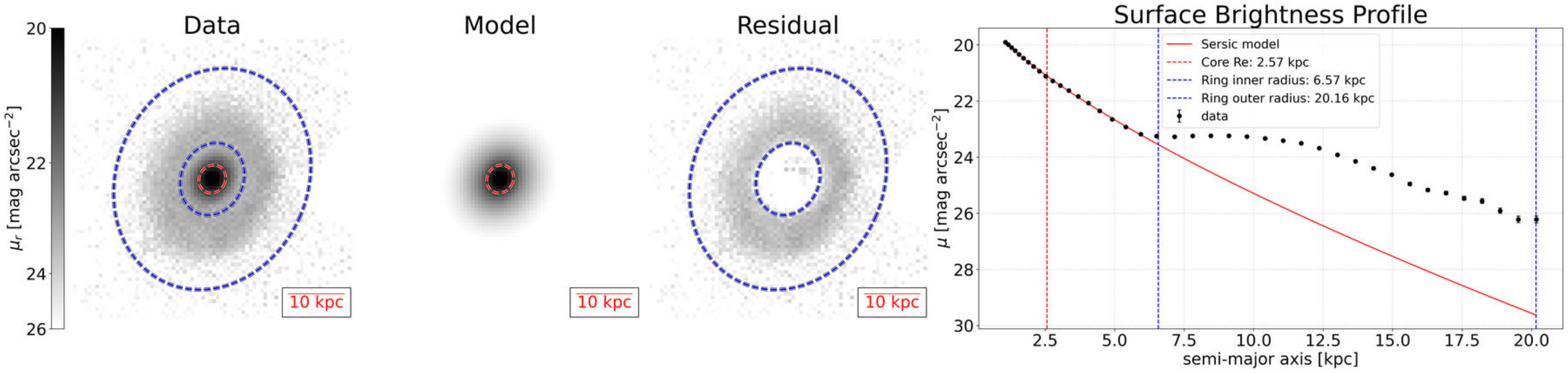}
    \atlaspanel{NGC 6028}{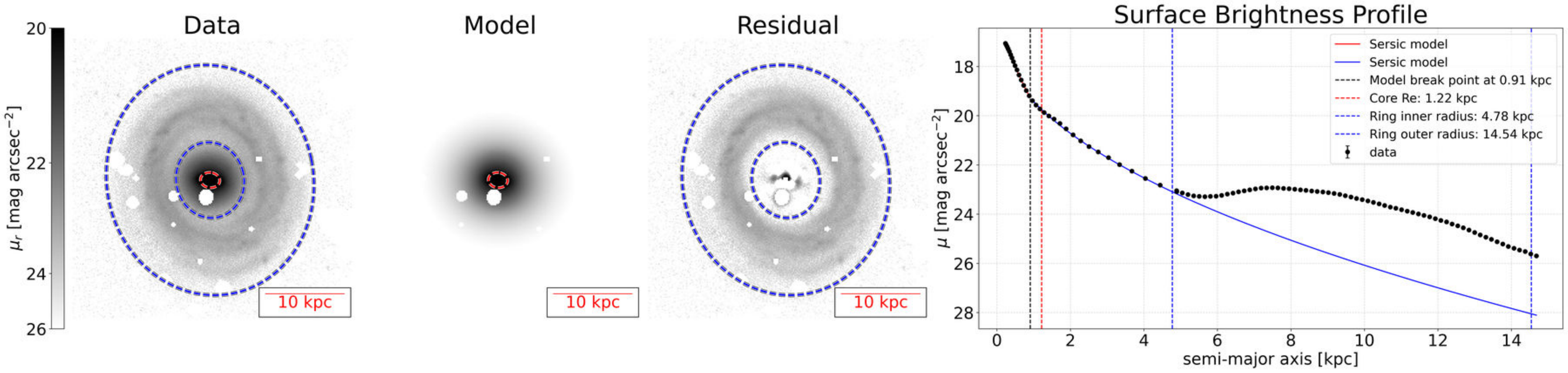}
    \atlaspanel{2MASX J01514310+0524439}{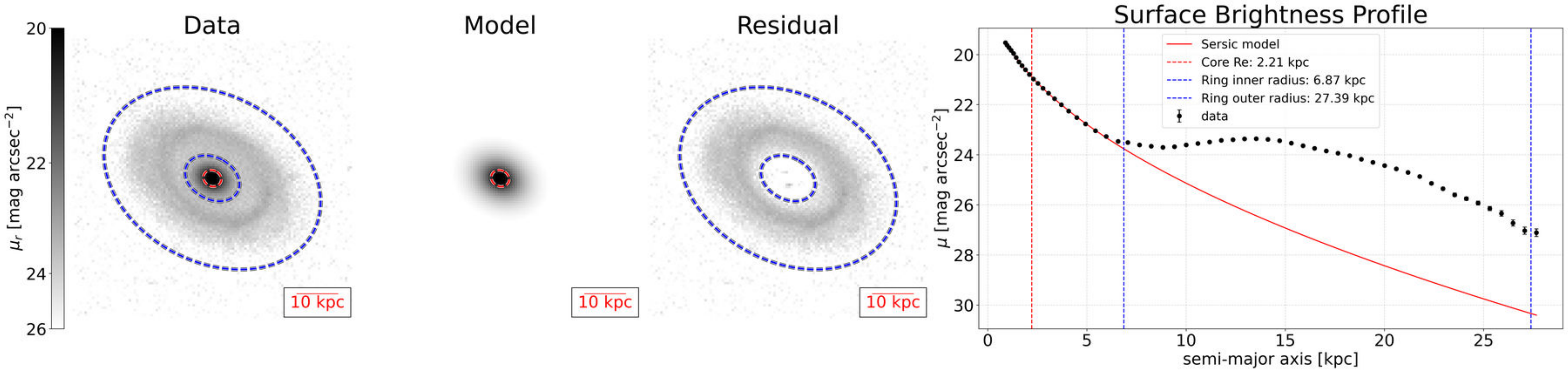}
    \atlaspanel{LEDA 3372664}{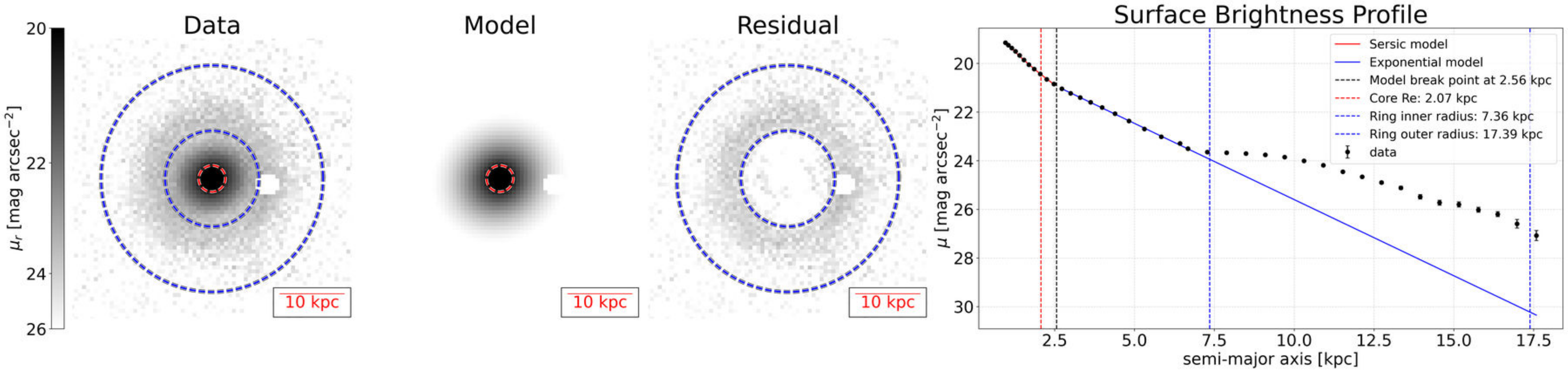}
    \atlaspanel{SDSS J215039.21+204249.0}{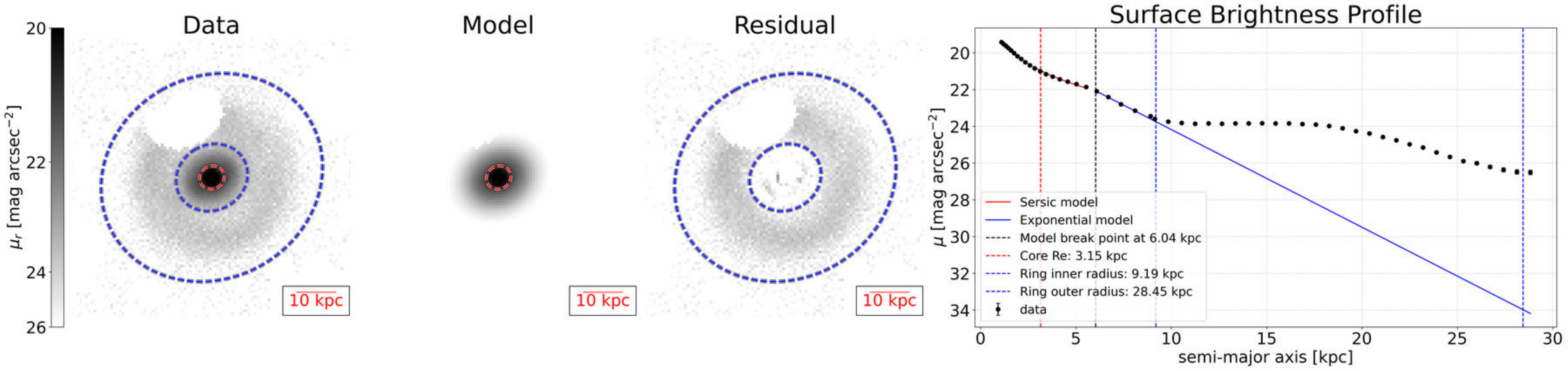}
    \caption{Individual galaxy decomposition figures (continued).}
    \label{fig:appendix_galaxy_atlas_6}
\end{figure*}

\clearpage
\let\atlaspanel\relax

\clearpage
\end{document}